\documentclass[12pt]{article}

\usepackage{geometry}
\usepackage{amsmath,amsthm}
\usepackage{amssymb}
\usepackage{booktabs}  
\usepackage{caption}   
\usepackage{enumitem}
\usepackage[labelformat=empty, position=top]{subcaption}
\usepackage[export]{adjustbox}
\usepackage{hyperref}
\usepackage{listings}
\usepackage{cite}
\usepackage{xcolor}
\definecolor{gray}{rgb}{0.4,0.4,0.4}
\definecolor{lightgray}{rgb}{.95,.95,.95}
\definecolor{darkblue}{rgb}{0.0,0.0,0.6}
\definecolor{cyan}{rgb}{0.0,0.6,0.6}

\theoremstyle{thmstyleone}%
\theoremstyle{thmstyletwo}%

\theoremstyle{thmstylethree}%
\newtheorem{definition}{Definition}%

\def\univs{\mathbb{U}}
\newcommand{\univ}[1]{\univs_{\mathit{#1}}}

\title{OCEL (Object-Centric Event Log) 2.0 Specification}
\date{\today}
\author{Alessandro Berti, Istv\'an Koren, Jan Niklas Adams, \\ Gyunam Park, Benedikt Knopp, Nina Graves, Majid Rafiei, Lukas Liß, \\ Leah Tacke Genannt Unterberg, Yisong Zhang, Christopher Schwanen, Marco Pegoraro, Wil M.P. van der Aalst}

\def\version{2.0}

\begin{document}


\makeatletter  
\begin{titlepage}
    \begin{center}
        {\Large \bfseries \@title}\\[1em]
        {\@author}\\[1em]
        {Chair of Process and Data Science, RWTH Aachen University}\\[1em]
        {\href{mailto:ocel@pads.rwth-aachen.de}{ocel@pads.rwth-aachen.de}}\\[1em]
        \hrule
    \end{center}
    \begin{flushleft}
        \textbf{Version:} \version\\
        \textbf{Date:} October 16, 2023\\ 
        \textbf{Standard Document URL:} \href{https://www.ocel-standard.org/2.0/ocel20_specification.pdf}{https://www.ocel-standard.org/2.0/ocel20\_specification.pdf}\\
        \textbf{Validation Schemes:} \\
        \begin{itemize}[noitemsep,nolistsep]
            \item XML: \href{https://www.ocel-standard.org/2.0/ocel20-schema-xml.xsd}{https://www.ocel-standard.org/2.0/ocel20-schema-xml.xsd}
            \item JSON: \href{https://www.ocel-standard.org/2.0/ocel20-schema-json.json}{https://www.ocel-standard.org/2.0/ocel20-schema-json.json}
            \item Relational: \href{https://www.ocel-standard.org/2.0/ocel20-schema-relational.pdf}{https://www.ocel-standard.org/2.0/ocel20-schema-relational.pdf}
        \end{itemize}
    \end{flushleft}
    \hrule

\vspace{1cm}

\begin{abstract}
Object-Centric Event Logs (OCELs) form the basis for Object-Centric Process Mining (OCPM).
OCEL 1.0 was first released in 2020 and triggered the development of a range of OCPM techniques.
OCEL 2.0 forms the new, more expressive standard, allowing for more extensive process analyses while remaining in an easily exchangeable format.
In contrast to the first OCEL standard, it can depict changes in objects, provide information on object relationships, and qualify these relationships to other objects or specific events.
Compared to XES, it is more expressive, less complicated, and better readable.
OCEL 2.0 offers three exchange formats: a relational database (SQLite), XML, and JSON format.
This OCEL 2.0 specification document provides an introduction to the standard, its metamodel, and its exchange formats, aimed at practitioners and researchers alike.
\end{abstract}

\end{titlepage}

\newpage
\tableofcontents
\newpage


\section{Scope and Structure of this Document}
\label{sec:scopeStructure}

This document introduces the OCEL 2.0 standard to record and exchange object-centric event logs.
The purpose of the standard is to guide the implementation of conformant process mining tools, and to provide the basis for the development of training material and other resources for users.

OCEL 2.0 and its metamodel are designed from the ground up to facilitate the exchange of event logs coming from a wide variety of information systems. Unlike traditional exchange formats, events may refer to any number of objects of different types.
It is intended to be interoperable with data extracted from a wide variety of databases, systems, or applications.
Likewise, the format aims to be equally compatible with existing and emerging Object-Centric Process Mining (OCPM) techniques.
It is anticipated that OCEL 2.0 will become the default exchange format for OCPM tools, whether these are research prototypes or commercial tools.

This document is structured as follows.
Section \ref{sec:introductionOCPM} explains object-centric process mining and discusses the limitations of current standards for recording object-centric event logs.
Section \ref{sec:introductionOCEL} introduces the OCEL 2.0 standard and discusses its advantages over past object-centric standardization attempts.
Section \ref{sec:formalDefinitions} contains the formal definitions of the standard.
Section \ref{sec:runningExample} illustrates OCEL 2.0 using a running example.
Sections \ref{sec:relationalImplementation}, \ref{sec:xmlImplementation} and \ref{sec:jsonImplementation} describe the practical implementation of the standard, using relational, XML, and JSON formats, respectively.
Finally, Section~\ref{sec:conclusion} concludes this document.

\section{Introduction to Object-Centric Process Mining}
\label{sec:introductionOCPM}

The first process mining algorithms were developed in the late 1990-ties \cite{process-mining-book-2016,PMhandbook-SS22}.
Initially, adoption was limited, with just a handful of researchers working on the topic. 
However, over time, the field matured. 
Currently, there are over 40 vendors offering process mining software (cf.\ \url{www.processmining.org})
and advisory firms such as Gartner consider these to form a new and substantial category of tools \cite{gartner-MQ-PM-2023}. 
Many of the world's largest companies already use process mining to improve their operational processes (across all economic sectors), and adoption is expected to increase in the coming years.
The increasing maturity of the process-mining discipline is also reflected by the success of the International Conference on Process Mining (ICPM) and the large number of process-mining papers in other conferences (e.g., BPM and CAiSE) and journals.

However, traditional process mining considers processes involving single cases, their events, and event attributes.
The approach falls short when dealing with complex, multi-dimensional processes, where events possibly relate to a variety of entities or \textit{objects} that interact and evolve over time \cite{mathematics-OCPM-wvda-2023}.
Traditional event data are based on the assumption that each event refers to precisely one case. The same applies to mainstream process modeling notations like Directly-Follows-Graphs (DFGs), BPMN models, UML activity diagrams, workflow nets, and process trees. However, most real-life events involve multiple objects. Traditional process mining approaches require the \textit{flattening} of event data in order to satisfy this assumption. This may lead to misleading analysis results. Process mining results also tend to become more complex because different objects get intertwined while trying to straitjacket the processes. Changing the viewpoint (e.g., looking at the process from a different angle) also implies changing the case notion
and going back to the source systems to extract other event data. This leads to redundancies in event data and unnecessary repetitions. Moreover, many compliance and performance problems arise at the intersection points of processes, systems, and organizations.

\emph{Object-Centric Process Mining} (OCPM) represents a paradigm shift, intended to address and overcome the inherent limitations of traditional case-centric process mining methods \cite{mathematics-OCPM-wvda-2023}. OCPM starts from the actual events and objects 
that leave traces in ERP (Enterprise Resource Planning), CRM (Customer Relationship Management), MES (Manufacturing Execution System), and other IT systems. In the databases of such systems, one-to-one relationships are the exception. Most relationships are one-to-many or many-to-many. As a result, data need to be transformed to be able to assign events to a single case, leading to all the problems mentioned before. Therefore, there is consensus among experienced process miners that \textit{Object-Centric Event Data} (OCED) provide a much better abstraction of reality than the classical case-based event logs. OCEL 1.0, released in 2020 \cite{OCEL-standard-2020-people}, was the first standard for storing OCED and triggered the development of OCPM techniques (e.g., discovering object-centric process models). OCEL 2.0 extends OCEL 1.0, leveraging experiences gathered while developing and applying these OCPM techniques.

Before discussing in what way OCEL 2.0 extends OCEL 1.0, we first need to introduce some terminology and basic concepts.
\begin{itemize}
\item \emph{Events}: Object-centric process mining works on discrete events. They represent the various actions or activities that occur within a system or process, such as approving an order, shipping an item, or making a payment. Every event is unique and corresponds to a specific action or observation at a specific point in time. Events are atomic (i.e., do not take time), have a timestamp, and may have additional attributes. Events are typed. 
\item \emph{Event Types}: Events are categorized into different types based on their nature or function. For example, a procurement process might have event types such as Order Created, Order Approved, or Invoice Sent. Each type of event represents a specific kind of action that can take place in the process. Each event is of exactly one type. Sometimes, we use the term \textit{activity} to refer to an event type. 
\item \emph{Objects}: In object-centric process mining, objects represent the entities that are involved in events. These might be physical items like products in a supply chain, machines, workers, or abstract/information entities like orders, invoices, or contracts in a procurement process. Objects have attributes with values, e.g., prices. These values may change over time.
\item \emph{Object Types}: Each object is of one type. The object is an instantiation of its type. Object types might include categories like Product, Order, Invoice, or Supplier.
\end{itemize}

Events and objects may be related. In particular, OCPM techniques exploit the following two relationships.
\begin{itemize}
\item \emph{Event-to-Object (E2O) Relationships}: Events are associated with objects. This relationship describes that an object affects an event or that an event affects an object.
In contrast to traditional event logs, events can be related to multiple objects. Furthermore, these relationships can be qualified differently, describing the role an object plays in the occurrence of this specific event.
Consider, for example, a meeting event involving multiple participant objects. Using a qualifier, it is possible to distinguish between regular participants and the organizer of the meeting.
\item \emph{Object-to-Object (O2O) Relationships}: Objects can also be related to other objects outside the context of an event.
For example, an employee may be part of an organizational unit.
In addition to the mere existence of a relation, this relationship can also be qualified (e.g., part-of, reports-to, or belongs-to).
\end{itemize}
Recent standardization attempts have addressed some but not all of these requirements. 
The first OCEL format (OCEL 1.0) provided an event log standard that could capture events related to multiple objects with attributes but did not include Object-to-Object (O2O) relationship, qualifiers for either O2O and E2O relationships, or changing object attribute values~\cite{OCEL-standard-2020-people,OCEL-ADBIS2021}. 
OCEL 2.0 addresses these limitations by providing a new metamodel and three storage formats, 
including a relational implementation of the standard. 
We will address OCEL 1.0, its limitations, and how OCEL 2.0 enriches the metamodel of OCEL 1.0 in more detail in the following section.

\section{Metamodel of the OCEL 2.0 Standard}
\label{sec:introductionOCEL}

Standards for storing object-centric event data serve as a crucial backbone in managing and analyzing complex process data. They provide a coherent, uniform, and structured approach to representing, storing, and exchanging event data across multiple systems, platforms, and applications. The adoption of a standard has several significant benefits:

\begin{itemize}
\item \emph{Interoperability}: a standard promotes seamless data interchange between diverse systems. It eliminates data silos by ensuring that event logs are represented in a universally understandable format.
\item \emph{Scalability}: a well-structured standard allows efficient handling of data, enabling it to scale with increasing complexity and volume. It ensures that the data remains manageable, reducing the overhead of dealing with unstructured or inconsistently structured logs.
\item \emph{Data Integrity and Consistency}: the standardization of event logs upholds the consistency and integrity of data across different sources. It provides a uniform structure to data, making it less prone to inconsistencies and errors, thereby improving the overall data quality.
\item \emph{Simplifies Analysis}: by adhering to a standard, the interpretation and analysis of event logs are significantly simplified. It enables the use of standard analysis tools and methods, fostering easy comparability and benchmarking of results.
\item \emph{Future-Proofing}: standards also future-proof data, ensuring that it remains accessible, reusable, and comprehensible even as technologies evolve.
\end{itemize}

The first comprehensive standard for storing event data was the IEEE Standard for eXtensible Event Stream (XES) \cite{XES-standard-2010}. XES became an official IEEE standard in 2016 \cite{ieee-CIM-XES-2017}. The revised standard (IEEE 1849-2023) was published on 8 September 2023 and will be valid for another ten years \cite{XES-standard-2023}.
XES has played a major role in the development of the field. 
However, within the process mining community, there seems to be a consensus that a paradigm shift is needed.
The development and adoption of a standard for storing object-centric event data are vital for realizing the full potential of process mining and other data-driven analytics methods. It paves the way for more effective, efficient, and reliable data management and analysis strategies.

The first version of the object-centric event log standard, OCEL 1.0, was a big step forward for object-centric process mining \cite{OCEL-standard-2020-people,OCEL-ADBIS2021}. It can store various types of events and objects in one log and link objects of different types to each event, giving a more detailed picture. OCEL 1.0 also allowed for adding multiple attributes to each event and object, providing even more information. This made data analysis deeper and more insightful. We provided OCEL 1.0 specifications in both JSON and XML formats. Several OCEL 1.0 data sets were provided and the availability of the standard 
fueled the development of a range of OCPM techniques, e.g.,
discovering object-centric Petri nets, discovering object-centric DFGs, checking conformance on object-centric process models, clustering object-centric event data, object-centric predictive methods, etc. \cite{mathematics-OCPM-wvda-2023,niklas-ocel-framework-icsoc-2022,ocpn_fi_2020}.

Although OCEL 1.0 can be considered a success, the time has come to extend the standard. 
OCEL 1.0 has a few deliberate limitations. 
In 2020, there were hardly any OCPM techniques, and the goal was to keep the standard as simple and lean as possible.
However, with the rapid development of the field, the first OCEL standard can now be perceived as an incomplete solution for object-centric process mining. In 2021, a survey was conducted by the IEEE Task Force on Process Mining \cite{XES-survey-icpm-ws-lnbip-2021}.
The goal was to collect requirements for a new standard succeeding XES. 
The online survey with 289 participants, spanning the roles of practitioners, researchers, software vendors and end-users,
showed the need for supporting object-centricity. 
This resulted in the so-called ``OCED Working Group'' of the IEEE Task Force on Process Mining. 
Input for the discussion was an early version of the OCEL 2.0 metamodel (similar to the model in \cite{mathematics-OCPM-wvda-2023}).
Unfortunately, the discussions in the OCED Working Group did not converge after 1.5 years of discussion. 
This was due to conflicting requirements (expressiveness versus simplicity), different 
implementation paradigms (relational versus graph-based), 
and a lack of clarity on who would implement things.
Therefore, after a delay of two years, the OCEL team decided to release OCEL 2.0, including 
example event logs, reference implementations, libraries, and documentation.
OCEL 2.0 aims to strike a middle ground between simplicity and expressiveness.
\begin{figure}[ht]
\includegraphics[width=\textwidth]{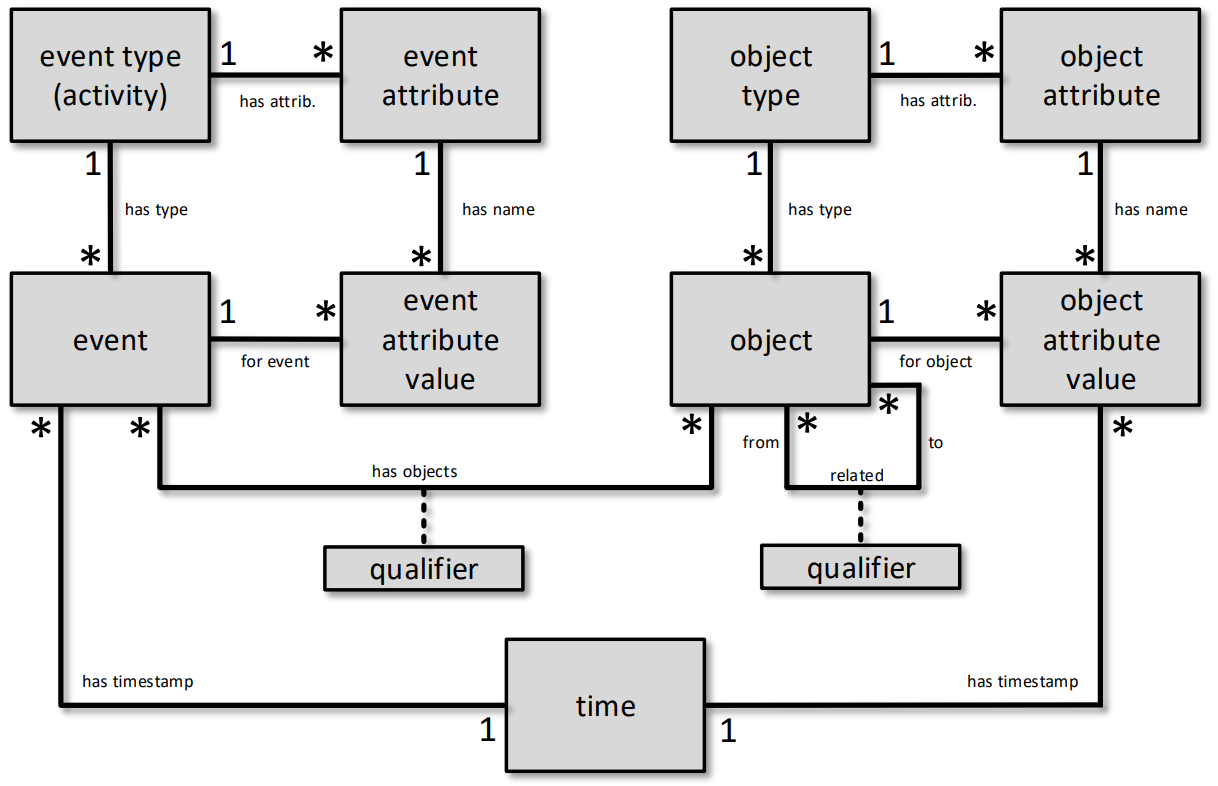}
\caption{OCEL 2.0 metamodel}
\label{fig:ocel20Metamodel}
\end{figure}

Figure~\ref{fig:ocel20Metamodel} shows the meta model using a simplified UML-like notation (just using classes, associations, and multiplicities). 
Compared to OCEL 1.0 there are several commonalities. There are objects and events, and these are typed.
Events have a timestamp and any number of additional attributes.
Also, objects can have attributes. There is a many-to-many relationship between events and objects.
There are also several differences. OCEL 2.0 was extended to address the limitations discussed before.

\begin{itemize}
\item \emph{Object-to-Object (O2O) Relationships}: OCEL 2.0 allows a deeper understanding of how objects interact within a business process. It shows that objects are part of a complex network of relationships and actions. Capturing these relationships can reveal insights about process performance and inefficiencies and allows for advanced analytics techniques like network analysis and predictive modeling.
\item \emph{Dynamic Object Attribute Values}: OCEL 2.0 adopts a dynamic approach where attribute values can change over time. Instead of having a single, fixed value, an object attribute may have a value that changes during the process. This gives a more realistic view of process instances by recognizing that object attributes change over time due to events and progression.
\item \emph{Relationship Qualifiers}: OCEL 2.0 offers capabilities to express qualifiers for relationships, both for Object-to-Object (O2O) and Event-to-Object (E2O) relationships. E2O relationship qualifiers describe in which role an object takes part in an event, while O2O relationship qualifiers can further characterize the association between two objects.
\end{itemize}

Apart from these conceptual extensions, the specifications have been enhanced. This makes the standard more scalable and easier to use in practical situations (e.g., in the context of a relational database). Notable differences are:
\begin{itemize}
\item \emph{Relational Specification based on Dense Tables}: One major feature of OCEL 2.0 is its data structure using dense tables. Each table corresponds to a unique event or object type, storing only relevant attributes. This results in efficient use of storage space and less data redundancy. It also scales well, allowing for easy addition of new event or object types. The separate tables make the data more accessible and easy to understand, improving both efficiency and analysis in process mining.
\item \emph{Improved XML Specification}: The XML specification in OCEL 2.0 has been significantly upgraded to handle complex data better. Essential information for events and objects is now directly within the corresponding tags, making the data more readable. It also includes the ability to show object-to-object relationships and track attribute changes over time, providing a better view of how objects evolve.
\end{itemize}

\section{Formal Definitions}
\label{sec:formalDefinitions}

The metamodel Figure~\ref{fig:ocel20Metamodel} is supported by a formalization that adds more details.  The theoretical foundation is crucial for understanding and using OCEL 2.0. These definitions form the basis for concrete exchange formats discussed later. The connection between theory and practice ensures that both the relational model and XML schema respect the standard's principles, enhancing its usefulness for object-centric event logging. Readers will see how these concepts turn into practical solutions, improving their understanding and use of OCEL 2.0 in process mining. We also encourage authors writing scientific papers using OCEL 2.0 to adopt these formal definitions and thus improve reliability. 

First, Definition~\ref{def:universes} introduces some concepts (universes) needed in the remainder.

\begin{definition}[Universes]\label{def:universes}
Let $\univ{\Sigma}$ be the universe of strings. We define the following pairwise disjoint universes:
\begin{itemize}
\item $\univ{ev} \subseteq \univ{\Sigma}$ is the universe of events.
\item $\univ{etype}  \subseteq \univ{\Sigma}$ is the universe of event types (i.e., activities).
\item $\univ{obj}  \subseteq \univ{\Sigma}$ is the universe of objects.
\item $\univ{otype}  \subseteq \univ{\Sigma}$ is the universe of object types.
\item $\univ{attr}  \subseteq \univ{\Sigma}$ is the universe of attribute names.
\item $\univ{val}$ is the universe of attribute values.
\item $\univ{time}$ is the universe of timestamps (with $0 \in \univ{time}$ as the smallest element and $\infty \in \univ{time}$ as the largest element)
\item $\univ{qual}  \subseteq \univ{\Sigma}$ is the universe of qualifiers.
\end{itemize}
\end{definition}

Note that the universes are assumed to be pairwise disjoint, i.e., objects cannot be used as events, etc.
$e \in \univ{ev}$ will be used to denote an event, $et \in \univ{etype}$ will be used to denote an event type,
$o \in \univ{obj}$ will be used to denote an object, $ot \in \univ{otype}$ will be used to denote an object type,
$ea \in \univ{attr}$ will be used to denote an event attribute, $oa \in \univ{attr}$ will be used to denote an object attribute,
$v \in \univ{val}$ will be used to denote an attribute value, $t \in \univ{time}$ will be used to denote a timestamp,
and $q \in \univ{qual}$ will be used to denote a qualifier.

We assume a total ordering on timestamps, with $0 \in \univ{time}$ as the earliest timestamp and $\infty \in \univ{time}$ as the latest
timestamp (i.e., for any $t \in \univ{time}$: $0 \leq t \leq \infty$).
These are added for notational convenience, e.g., 
we can use $0$ for missing timestamps and the start of the process,
and $\infty$ as the end time.
We would like to emphasize that these two reference timestamps (i.e., $0$ and $\infty$)
are chosen for convenience. In the formalization, time is mapped on the non-negative reals, 
but concrete implementations will use, for example, the ISO 8601 time format.

Definition~\ref{def:ocelDef} provides the formal definition for object-centric event logs, describing all the basic concepts introduced in Section~\ref{sec:introductionOCPM} (events, objects, event types, object types, and event-to-object relationships), and introducing object-to-object relationships and dynamic object attribute values.

\begin{definition}[OCEL]\label{def:ocelDef}
An Object-Centric Event Log (OCEL) is a tuple \( L = (E, O, EA, \)
\\
\(  OA, \textrm{evtype}, \textrm{time}, \textrm{objtype}, \textrm{eatype}, \textrm{oatype}, \textrm{eaval}, \textrm{oaval}, \textrm{E2O}, \textrm{O2O}) \) with
\begin{itemize}
\item $E \subseteq \univ{ev}$ is the set of events.
\item $O \subseteq \univ{obj}$ is the set of objects.
\item $\textrm{evtype} : E \rightarrow \univ{etype}$ assigns types to events.
\item $\textrm{time} : E \rightarrow \univ{time}$ assigns timestamps to events.
\item $\textrm{EA} \subseteq \univ{attr}$ is the set of event attributes.
\item $\textrm{eatype} : EA \rightarrow \univ{etype}$ assigns event types to event attributes.
\item $\textrm{eaval} : (E \times EA) \not\rightarrow \univ{val}$ assigns values to event attributes (not all the attributes are mapped for each event).
\item $\textrm{objtype} : O \rightarrow \univ{otype}$ assigns types to objects.
\item $OA \subseteq \univ{attr}$ is the set of object attributes.
\item $\textrm{oatype} : OA \rightarrow \univ{otype}$ assigns object types to object attributes.
\item $\textrm{oaval} : (O \times OA \times \univ{time}) \not\rightarrow \univ{val}$ assigns values to object attributes.
\item $E2O \subseteq E \times \univ{qual} \times O$ are the qualified event-to-object relations.
\item $O2O \subseteq O \times \univ{qual} \times O$ are the qualified object-to-object relations.
\end{itemize}
such that
\begin{itemize}
\item $\textrm{dom}(\textrm{eaval}) \subseteq \{ (e, ea) \in E \times EA ~ \arrowvert ~ \textrm{evtype}(e) = \textrm{eatype}(ea) \}$ to ensure that only existing event attributes can have values.
\item $\textrm{dom}(\textrm{oaval}) \subseteq \{ (o, oa, t) \in O \times OA \times \univ{time} ~ \arrowvert ~ \textrm{objtype}(o) = \textrm{oatype}(oa) \}$ to ensure that only
existing object attributes can have values.
\end{itemize}
\end{definition}

The stipulations embedded in the final two criteria of the definition ensure that each event and object is limited to possessing attribute values pertinent to its respective event or object type. Additionally, these guidelines also mandate that the set of attributes is distinct and non-overlapping for each individual event and object type, guaranteeing a disjoint attribute set across all types.

In order to facilitate the following examples and explanations, we introduce the following notations given an OCEL $L$:
\begin{itemize}
\item $ET(L) = \{ \textrm{evtype}(e) ~ \arrowvert ~ e \in E \}$ is the set of event types.
\item $OT(L) = \{ \textrm{objtype}(o) ~ \arrowvert ~ o \in O \}$ is the set of object types.
\item For any event $e \in E$ and event attribute $\textrm{ea} \in \univ{attr}$:
\begin{itemize}
\item $\textrm{eaval}_{ea}(e) = \textrm{eaval}(e, ea)$ if $(e, ea) \in \textrm{dom}(\textrm{eaval})$.
\item $\textrm{eaval}_{ea}(e) = \bot$ if $(e, ea) \not\in \textrm{dom}(\textrm{eaval})$.
\end{itemize}
\item For any object $o \in O$, object attribute $\textrm{oa} \in \univ{attr}$ and time $t \in \univ{time}$:
\begin{itemize}
\item $\textrm{oaval}_{oa}^{t}(o) = \textrm{oaval}(o, oa, t')$ if there exists a $t' \in \univ{time}$ such that $t' \leq t$ and $(o, oa, t') \in \textrm{dom}(\textrm{oaval})$
such that there is no $t'' \in \univ{time}$ such that $t' < t'' \leq t$ and $(o, oa, t'') \in \textrm{dom}(\textrm{oaval})$.
\item If no such $t'$ exists, then $\textrm{oaval}_{oa}^{t}(o) = \bot$.
\end{itemize}
Hence, $\textrm{oaval}_{oa}^{t}$ provides us with the latest object attribute value at time $t$.
\item $\textrm{oaval}_{oa}(o) = \textrm{oaval}_{oa}^{\infty}(o)$ is the final value for the object attribute in the event log.
\end{itemize}

Note that $\textrm{oaval}$ describes object attribute updates at particular points in time. 
Function $\textrm{oaval}_{oa}^{t}(o)$ allows us to determine the value of object attributes at any point in time, thus clarifying the semantics.
A missing timestamp for a value assignment can be interpreted as time $0$.
If this is the only value assignment for an object-attribute combination, then the value for an object attribute is always the same. This way, we can handle static object-attribute values.

\begin{figure}[ht]
\includegraphics[width=\textwidth]{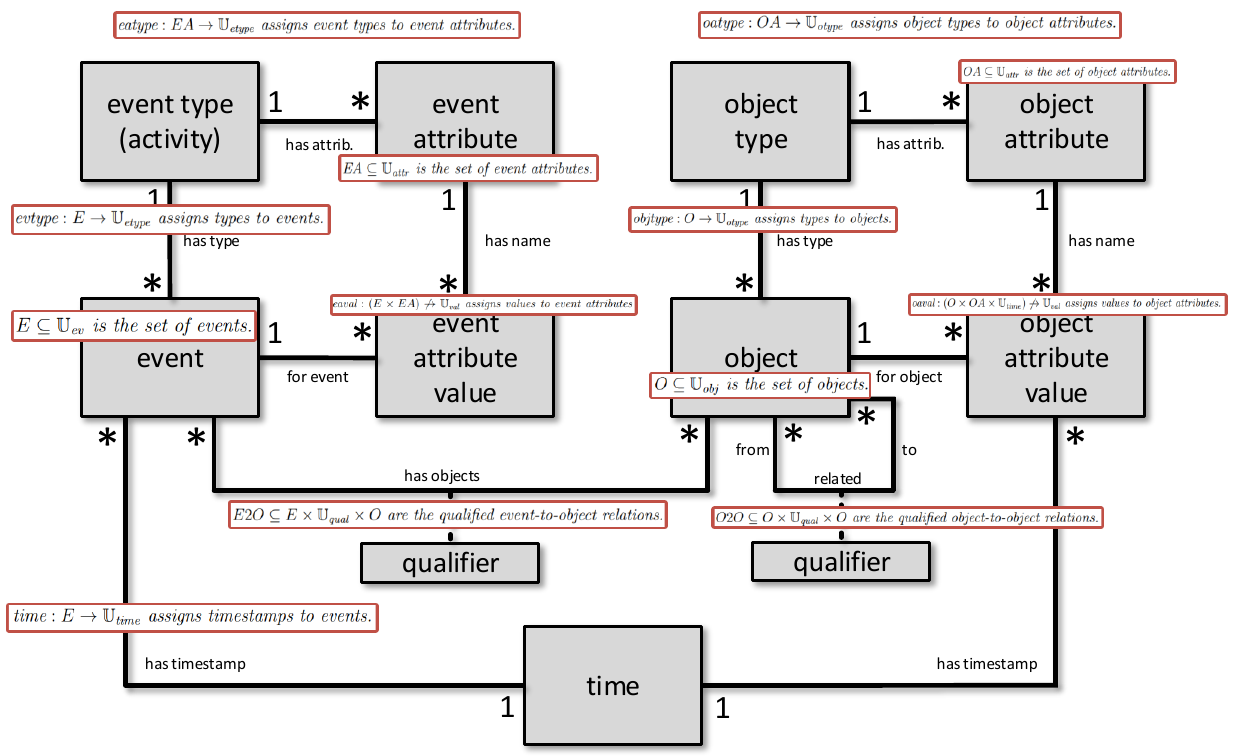}
\caption{Linking the OCEL 2.0 metamodel to the formalization in Definition~\ref{def:ocelDef}.}
\label{fig:ocel20MetamoedlFig2}
\end{figure}

Object attribute values are deliberately \textit{not} connected to events. The definition allows for events without objects or objects without events. Events should correspond to relevant activities, and therefore, there should not be the need to promote individual object attribute changes to events. Also, the object-to-object relations may exist independent of events. The only explicit connections between events and objects are the event-to-object relations (i.e., \emph{E2O}). However, for an event $e$ happening at time $t$ involving object $o$ with attribute $\textrm{oa}$ we can look up the corresponding value at the time of the event via $\textrm{oaval}_{oa}^{t}(o)$. Hence, there is no limitation.

Figure~\ref{fig:ocel20MetamoedlFig2} links the OCEL metamodel in Figure~\ref{fig:ocel20Metamodel} to the formalization just provided.
Sections \ref{sec:relationalImplementation}, \ref{sec:xmlImplementation}, and \ref{sec:jsonImplementation} map the formalization onto concrete relational, XML, and JSON formats.

\section{Running Example}
\label{sec:runningExample}

In order to visualize the concepts of Definition~\ref{def:ocelDef} and propose some implementations of OCEL 2.0, we describe an example object-centric setting supporting the following situations:
\begin{enumerate}
\item A purchase requisition ({\bf PR1}) is created and approved, requesting $500$ cows, and a purchase order ({\bf PO1}) is created on top of the purchase requisition.
The quantity of the purchase order {\bf PO1} is then changed to $600$ cows. Two invoices {\bf R1} and {\bf R2} are received (for the first $500$ cows and for the additional $100$ cows),
which are paid separately by the payments {\bf P1} and {\bf P2}.
\item Mario is an unethical employee who places purchase orders of notebooks without getting proper approval. An invoice {\bf R3} is received before an order {\bf PO2} is formally inserted in the system. Therefore,
Sam, who is a financial controller, blocks the payment of the invoice. However, Mario manages to override Sam and puts himself as an approver of the invoice; therefore, the invoice is paid in the {\bf P4} payment.
\end{enumerate}

In this example, the attributes at the object level evolve, highlighting one of the main features of OCEL 2.0.
Also, we provide qualified event-to-object and object-to-object relationships, specifically in Table \ref{tab:tabE2O}
and Table \ref{tab:tabO2O}, highlighting the other main novelties.

For this example:
\begin{itemize}
    \item We identify the following sets (in relation to Definition~\ref{def:ocelDef}):
    \begin{itemize}
    \item (sets of objects) $O = \{ \textbf{PR1}, \textbf{PO1}, \textbf{R1}, \textbf{R2}, \textbf{P1}, \textbf{P2}, \textbf{R3}, \textbf{PO2}, \textbf{R3}, \textbf{P3} \}$.
    \item (sets of events) $E = \{ e1, e2, e3, e4, e5, e6, e7, e8, e9, e10, e11, e12, e13 \}$.
    \item (sets of attributes at the event level) $EA = \{ \textrm{pr\_creator}, \\ \textrm{pr\_approver}, \textrm{po\_creator}, \textrm{po\_editor}, \textrm{invoice\_inserter}, \textrm{invoice\_blocker}, \\ \textrm{invoice\_block\_rem}, \textrm{payment\_inserter} \}$ where:
    \begin{itemize}
        \item {\bf pr\_creator} is the resource that created the purchase requisition in the system.
        \item {\bf pr\_approver} is the resource that approved the purchase requisition in the system.
        \item {\bf po\_creator} is the resource that created the purchase order in the system.
        \item {\bf po\_editor} is the resource that changed some parameters of the purchase order.
        \item {\bf invoice\_inserter} is the resource that inserted the invoice in the system.
        \item {\bf invoice\_blocker} is the resource that blocked the payment of a given invoice.
        \item {\bf invoice\_block\_rem} is the resource that removed the payment block.
        \item {\bf payment\_inserter} is the resource that performed the payment.
    \end{itemize}
    \item (sets of attributes at the object level) \\ $OA = \{ \textrm{pr\_product}, \textrm{pr\_quantity}, \textrm{po\_product}, \textrm{po\_quantity}, \textrm{is\_blocked} \}$ where:
    \begin{itemize}
        \item {\bf pr\_product} is the product requested in the purchase requisition.
        \item {\bf pr\_quantity} is the quantity requested in the purchase requisition.
        \item {\bf po\_product} is the product bought in the purchase order.
        \item {\bf po\_quantity} is the quantity bought in the purchase order.
        \item {\bf is\_blocked} relates to a payment block on the invoice.
    \end{itemize}
    \end{itemize}
    \item The object types corresponding to the given objects ($\textrm{objtype}$ in Definition~\ref{def:ocelDef}) are described in Table~\ref{tab:highLevelOtAttrs}. Moreover, the attributes defined for the object types ($\textrm{oatype}$ in Definition~\ref{def:ocelDef}) are also described in Table~\ref{tab:highLevelOtAttrs}.
    \item The event types are described in Table~\ref{tab:highLevelEtAttrs}, along with the corresponding attributes ($\textrm{eatype}$ in Definition~\ref{def:ocelDef}) and event identifiers.
    \item A high-level tabular view is provided in Table~\ref{tab:tabularViewObjectCentricEventLog}. In particular, the event type ($\textrm{evtype}$ in Definition~\ref{def:ocelDef}) and timestamp ($\textrm{time}$ in Definition~\ref{def:ocelDef}) of each event is defined.
\end{itemize}

\begin{table}[ht]
    \centering
    \caption{High-level view on the object types of the object-centric event log (running example).}
    \begin{tabular}{lcl}
        \toprule
        \textbf{Object Type} & \textbf{Attributes} & \textbf{Object IDs} \\
        \midrule
        Purchase Requisition & pr\_product, pr\_quantity & \textbf{PR1} \\
        Purchase Order       & po\_product, po\_quantity & \textbf{PO1, PO2} \\
        Invoice              & is\_blocked              & \textbf{R1, R2, R3} \\
        Payment              & $\emptyset$              & \textbf{P1, P2} \\
        \bottomrule
    \end{tabular}
    \label{tab:highLevelOtAttrs}
\end{table}

\begin{table}[ht]
    \centering
    \caption{High-level view on the event types of the object-centric event log (running example).}
    \begin{tabular}{lcl}
        \toprule
        \textbf{Event Type} & \textbf{Attributes} & \textbf{Event IDs} \\
        \midrule
        Create Purchase Requisition & pr\_creator & e1 \\
        Approve Purchase Requisition & pr\_approver & e2 \\
        Create Purchase Order & po\_creator & e3, e10 \\
        Change PO Quantity & po\_editor & e4 \\
        Insert Invoice & invoice\_inserter & e5, e6, e9 \\
        Set Payment Block & invoice\_blocker & e11 \\
        Remove Payment Block & invoice\_block\_rem & e12 \\
        Insert Payment & payment\_inserter & e7, e8, e13 \\
        \bottomrule
    \end{tabular}
    \label{tab:highLevelEtAttrs}
\end{table}

\begin{table}
    \centering
    \caption{High-level tabular view on the object-centric event log (running example), showing the list of events along with the related objects. In this view, no event/object attribute is reported, no qualifier is reported (see Table \ref{tab:tabE2O} for the event-to-object relationships qualifiers), and no object-to-object relationship is described.}
    \begin{tabular}{lcll}
        \toprule
        \textbf{Event ID} & \textbf{Event Type} & \textbf{Timestamp} & \textbf{Related Objects} \\
        \midrule
        e1  & Create Purchase Requisition  & 2022-01-09 15:00 & \textbf{PR1} \\
        e2  & Approve Purchase Requisition  & 2022-01-09 16:30 & \textbf{PR1} \\
        e3  & Create Purchase Order         & 2022-01-10 09:15 & \textbf{PR1, PO1} \\
        e4  & Change PO Quantity            & 2022-01-13 12:00 & \textbf{PO1} \\
        e5  & Insert Invoice                & 2022-01-14 12:00 & \textbf{PO1, R1} \\
        e6  & Insert Invoice                & 2022-01-16 11:00 & \textbf{PO1, R2} \\
        e7  & Insert Payment                & 2022-01-30 23:00 & \textbf{R1, P1} \\
        e8  & Insert Payment                & 2022-01-31 22:00 & \textbf{R2, P2} \\
        e9  & Insert Invoice                & 2022-02-02 09:00 & \textbf{R3} \\
        e10 & Create Purchase Order         & 2022-02-02 17:00 & \textbf{R3, PO2} \\
        e11 & Set Payment Block             & 2022-02-03 07:30 & \textbf{R3} \\
        e12 & Remove Payment Block          & 2022-02-03 23:30 & \textbf{R3} \\
        e13 & Insert Payment                & 2022-02-28 23:00 & \textbf{R3, P3} \\
        \bottomrule
    \end{tabular}
    \label{tab:tabularViewObjectCentricEventLog}
\end{table}


\begin{figure*}[!t]
\centering
\includegraphics[width=0.6\textwidth]{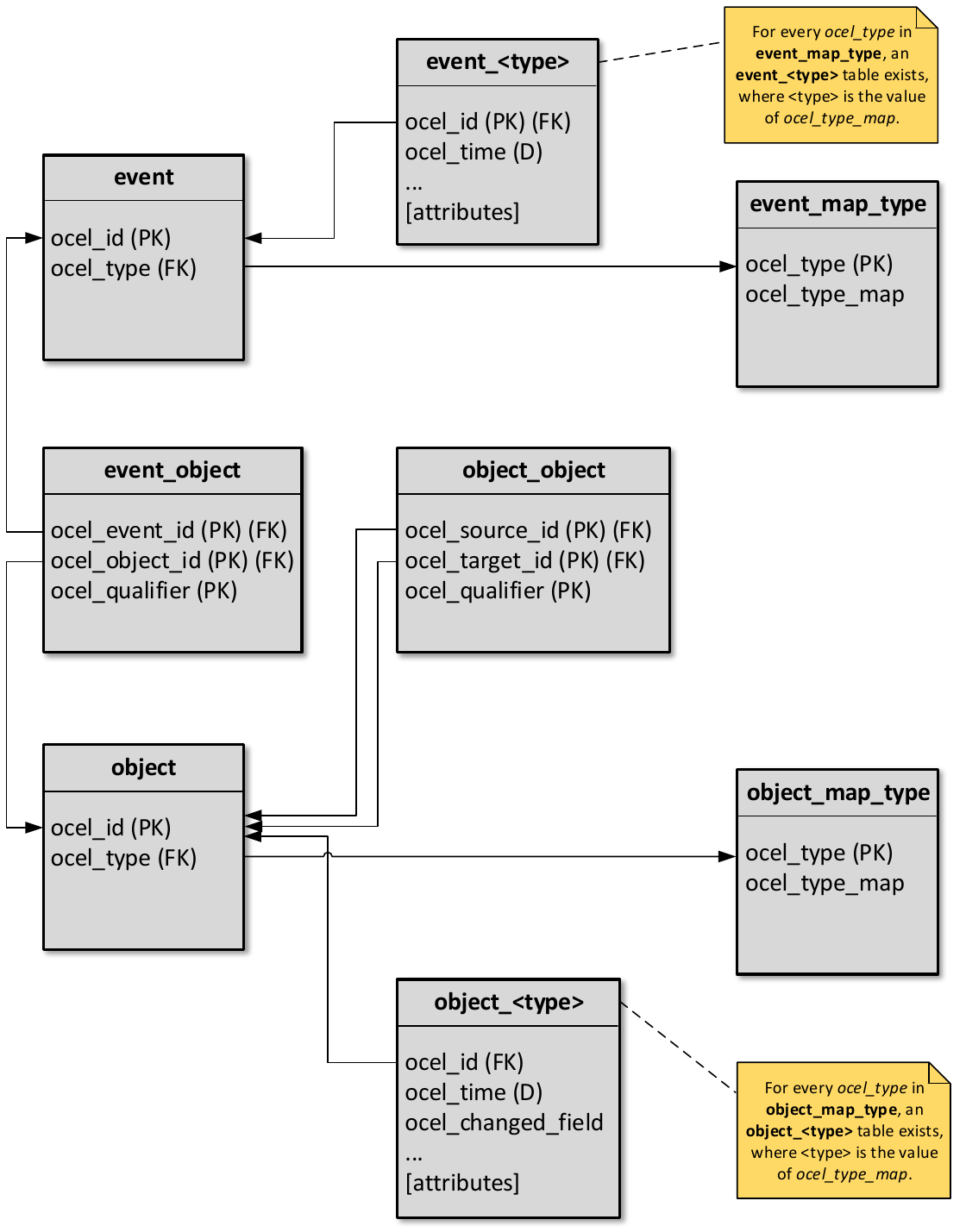}
\caption{General relational schema of the proposed relational implementation.}
\label{fig:generalSchemaProposedRelational}
\end{figure*}

\clearpage
\newpage

\section{Relational SQLite Format}
\label{sec:relationalImplementation}

\begin{figure*}
\centering
\includegraphics[height=24cm]{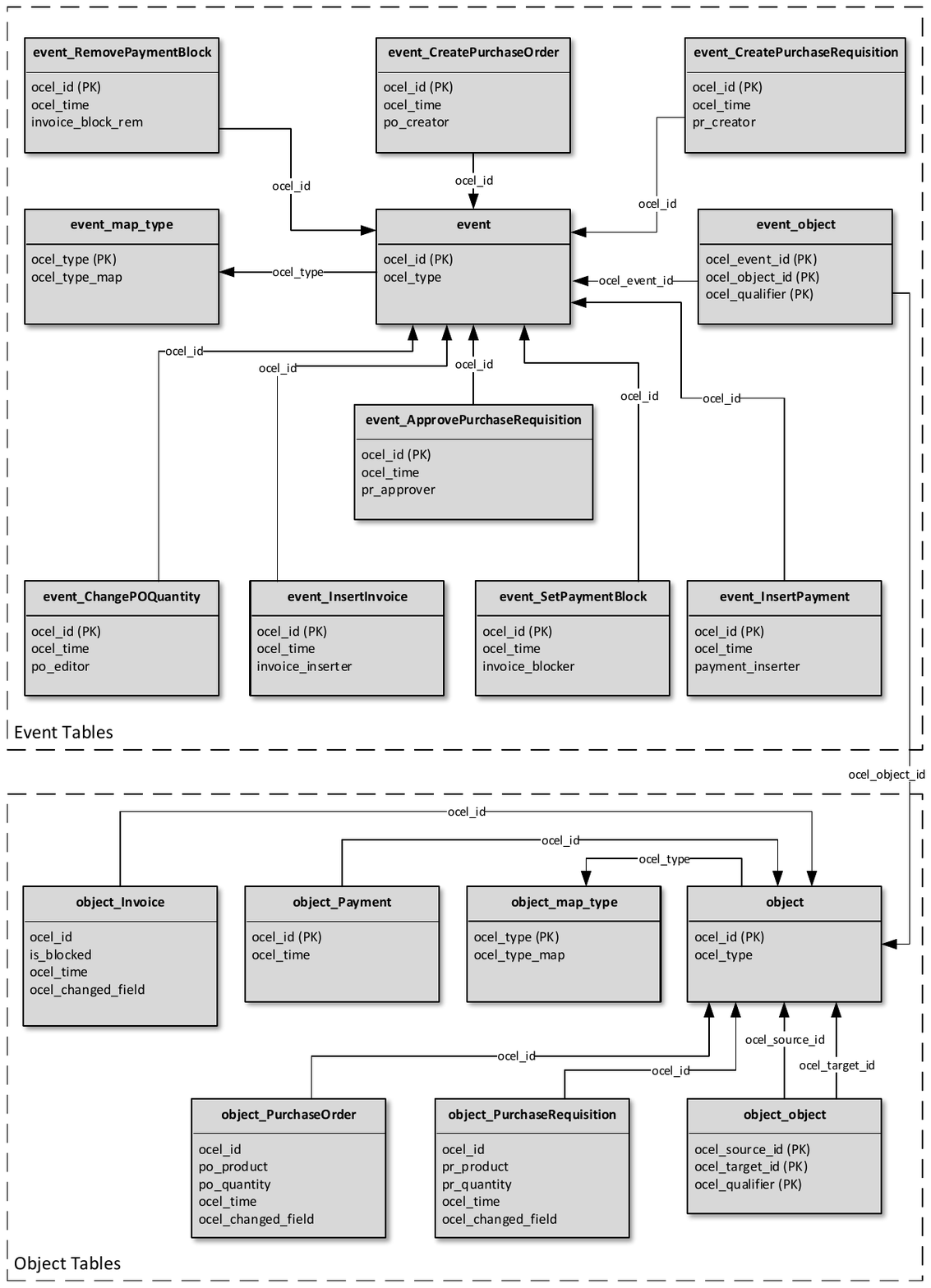}
\caption{Relational schema of the running example OCEL 2.0 log.}
\label{fig:relationalSchemaOCEL20}
\end{figure*}

We propose a relational implementation of the standard, 
which adheres to Definition~\ref{def:ocelDef}. In this implementation, starting from an object-centric event log $L$, we have:
\begin{itemize}
    \item We have a table, {\bf event\_map\_type}, reporting the distinct event types ($ET(L)$); and a table, {\bf object\_map\_type}, reporting the distinct object types ($OT(L)$).
    \item We have a table, {\bf event}, reporting the event type ($\textrm{evtype}$ in Definition~\ref{def:ocelDef}) for each event, and a table, {\bf object}, reporting the object type ($\textrm{objtype}$ in Definition~\ref{def:ocelDef}) for each object. This is described in Subsection \ref{subsec:eventsObjectsTables}.
    \item For every event type in $ET(L)$, we have a different table. More specifically, if $\textrm{et} \in ET(L)$ is the event type, we have a table called {\bf event\_$\oplus${$map_{ET}$}(\textrm{et})}, where $map_{ET}: ET(L) \rightarrow \univ{\Sigma}$ is an injective function\footnote{Which can be the identity function.} mapping the event types to unique identifiers, containing the events of the given event type along with their timestamp and attributes. This is described in Subsection \ref{subsec:eventTypeTables}.
    \item For every object type in $OT(L)$, we have a different table. More specifically, if $ot \in OT(L)$ is the object type, we have a table called {\bf object\_$\oplus${$map_{OT}$}(ot)}, where $map_{OT}: OT(L) \rightarrow \univ{\Sigma}$ is an injective function\footnote{Which can be the identity function.} mapping the object types to unique identifiers, containing the objects of the given object type along with the history of their attributes. This is described in Subsection \ref{subsec:objTypeTables}.
    \item We have a table, {\bf event\_object}, containing the event-to-object relationships ($\textrm{E2O}$ in Definition~\ref{def:ocelDef}).  This is described in Subsection \ref{subsec:evToObjRelationships}.
    \item We have a table, {\bf object\_object}, containing the object-to-object relationships ($\textrm{O2O}$ in Definition~\ref{def:ocelDef}). This is described in Subsection \ref{subsec:objObjRelationships}.
\end{itemize}

A general representation of the relational schema of OCEL 2.0 is portrayed in Figure~\ref{fig:generalSchemaProposedRelational}.
The overall relational schema of the running example OCEL 2.0 log is shown in Figure~\ref{fig:relationalSchemaOCEL20}.

\clearpage
\newpage

\subsection{Tables for the Distinct Event/Object Types}
\label{subsec:tablesDistinctEventObjTypes}

In our implementation, we define the table {\bf event\_map\_type} (Table~\ref{tab:distinctEventTypes}) defining the event types along with their unique identifier obtained by applying the injective function $map_{ET} : ET(L) \rightarrow \univ{\Sigma}$. This unique identifier is used to link the event type to a table containing the attributes of the events of the given event type (see Subsection \ref{subsec:eventTypeTables}). The event type should be set as the primary key to avoid duplication.

\begin{table}[ht]
\centering
\caption{[Proposed relational implementation] Distinct event types: \textbf{event\_map\_type}}
\begin{tabular}{ll}
\toprule
\textbf{ocel\_type [PK]} & \textbf{ocel\_type\_map} \\
\midrule
Approve Purchase Requisition & ApprovePurchaseRequisition \\
Change PO Quantity & ChangePOQuantity \\
Create Purchase Order & CreatePurchaseOrder \\
Create Purchase Requisition & CreatePurchaseRequisition \\
Insert Invoice & InsertInvoice \\
Insert Payment & InsertPayment \\
Remove Payment Block & RemovePaymentBlock \\
Set Payment Block & SetPaymentBlock \\
\bottomrule
\end{tabular}
\label{tab:distinctEventTypes}
\end{table}

Also, the table {\bf object\_map\_type} (Table~\ref{tab:distinctObjectTypes}) defines the object types along with their unique identifier obtained applying the injective function $map_{OT} : OT(L) \rightarrow \univ{\Sigma}$. This unique identifier is used to link the object type to a table containing the attributes of the objects of the given object type (see Subsection \ref{subsec:objTypeTables}). The object type should be set as the primary key to avoid duplication.

\begin{table}[ht]
\centering
\caption{Proposed relational implementation: Distinct object types}
\label{tab:distinctObjectTypes}
\begin{tabular}{ll}
\toprule
\textbf{ocel\_type [PK]} & \textbf{ocel\_type\_map} \\
\midrule
Invoice                 & Invoice \\
Payment                 & Payment \\
Purchase Order          & PurchaseOrder \\
Purchase Requisition    & PurchaseRequisition \\
\bottomrule
\end{tabular}
\end{table}

\clearpage
\newpage

\subsection{Events and Objects Tables}
\label{subsec:eventsObjectsTables}

In the proposed implementation, we use several tables to store information related to an event/object type. However, we also create two additional tables, hosting the event/object identifiers. This is done to map the event/object to a specific event/object type table, and to allow for the definition of E2O/O2O tables with proper foreign keys (directed towards the events and the objects tables). 

This is done in Table~\ref{tab:generalTabEvents} (having the event identifier, \emph{ocel\_id}, as primary key) and Table~\ref{tab:generalTabObjects} (having the object identifier, \emph{ocel\_id}, as primary key).

\begin{table}[ht]
    \centering
    \captionsetup{justification=centering}
    \caption{[Proposed Relational Implementation] General Events Table \\ \textbf{event}}
    \label{tab:generalTabEvents}
    \begin{tabular}{@{} l l @{}}
        \toprule
        \textbf{ocel\_id [PK]} & \textbf{ocel\_type FK} \\
        \midrule
        e1  & Create Purchase Requisition \\
        e2  & Approve Purchase Requisition \\
        e3  & Create Purchase Order \\
        e4  & Change PO Quantity \\
        e5  & Insert Invoice \\
        e6  & Insert Invoice \\
        e7  & Insert Payment \\
        e8  & Insert Payment \\
        e9  & Insert Invoice \\
        e10 & Create Purchase Order \\
        e11 & Set Payment Block \\
        e12 & Remove Payment Block \\
        e13 & Insert Payment \\
        \bottomrule
    \end{tabular}
\end{table}

\begin{table}[ht]
    \centering
    \captionsetup{justification=centering}
    \caption{[Proposed Relational Implementation] General Objects Table \\ \textbf{object}}
    \label{tab:generalTabObjects}
    \begin{tabular}{@{} l l @{}}
        \toprule
        \textbf{ocel\_id [PK]} & \textbf{ocel\_type FK} \\
        \midrule
        PR1 & Purchase Requisition \\
        PO1 & Purchase Order \\
        PO2 & Purchase Order \\
        R1  & Invoice \\
        R2  & Invoice \\
        R3  & Invoice \\
        P1  & Payment \\
        P2  & Payment \\
        \bottomrule
    \end{tabular}
\end{table}

\clearpage
\newpage

\subsection{Event Type Tables}
\label{subsec:eventTypeTables}

Having defined the table {\bf event} as the general table for the events (having the identifier of the event as the primary key), we define event-type-specific tables. These allow for the storage of event-type-specific attributes without having a gigantic ``sparse table'' hosting all the events with all the different attributes (with an empty value for most of them). Therefore, if $et \in ET(L)$ is the event type, we have a table called {\bf event\_$\oplus${$map_{ET}$}(et)}, where $map_{ET}: ET(L) \rightarrow \univ{\Sigma}$ is an injective function\footnote{Which can be the identity function.} mapping the event types to unique identifiers, containing the events of the given event type along with their timestamp and attributes. In particular, all the attributes associated with the given event type ($\textrm{eatype}$ in Definition~\ref{def:ocelDef}) are columns of the table, and the values for each event are populated accordingly ($\textrm{eaval}$ in Definition~\ref{def:ocelDef}). Moreover, the timestamp is a column of the event-type-specific tables. The event identifier column is pointing (as a foreign key) to the event identifier column of the {\bf event} table. Examples follow (Table~\ref{tab:tabEtCreatePR}, Table~\ref{tab:tabEtApprovePR}, Table~\ref{tab:tabEtCreatePO}, Table~\ref{tab:tabChangePOQuantity}, Table~\ref{tab:tabInsertInvoice}, Table~\ref{tab:tabSetPaymentBlock}, Table~\ref{tab:tabRemovePaymentBlock}, Table~\ref{tab:tabInsertPayment})  for all the event types of the running example log.

\begin{table}[ht]
    \centering
    \caption{[Proposed relational implementation] Event Type Table: \textbf{event\_CreatePurchaseRequisition}}
    \label{tab:tabEtCreatePR}
    \begin{tabular}{@{} l l l @{}}
        \toprule
        \textbf{ocel\_id [PK] [FK]} & \textbf{ocel\_time} & \textbf{pr\_creator} \\
        \midrule
        e1 & 2022-01-09 15:00 & Mike \\
        \bottomrule
    \end{tabular}
\end{table}

\begin{table}[ht]
  \centering
  \caption{[Proposed relational implementation] Event type table: \textbf{event\_ApprovePurchaseRequisition}}
  \begin{tabular}{@{}ccc@{}}
    \toprule
    \textbf{ocel\_id [PK] [FK]} & \textbf{ocel\_time} & \textbf{pr\_approver} \\
    \midrule
    e2 & 2022-01-09 16:30 & Tania \\
    \bottomrule
  \end{tabular}
  \label{tab:tabEtApprovePR}
\end{table}

\begin{table}[ht]
  \centering
  \caption{[Proposed relational implementation] Event type table: \textbf{event\_CreatePurchaseOrder}}
  \begin{tabular}{@{}ccc@{}}
    \toprule
    \textbf{ocel\_id [PK] [FK]} & \textbf{ocel\_time} & \textbf{po\_creator} \\
    \midrule
    e3  & 2022-01-10 09:15 & Mike  \\
    e10 & 2022-02-02 17:00 & Mario \\
    \bottomrule
  \end{tabular}
  \label{tab:tabEtCreatePO}
\end{table}

\begin{table}[ht]
  \centering
  \caption{[Proposed relational implementation] Event type table: \textbf{event\_ChangePOQuantity}}
  \begin{tabular}{@{}ccc@{}}
    \toprule
    \textbf{ocel\_id [PK] [FK]} & \textbf{ocel\_time} & \textbf{po\_editor} \\
    \midrule
    e4 & 2022-01-13 12:00 & Mike \\
    \bottomrule
  \end{tabular}
  \label{tab:tabChangePOQuantity}
\end{table}

\begin{table}[ht]
  \centering
  \caption{[Proposed relational implementation] Event type table: \textbf{event\_InsertInvoice}}
  \begin{tabular}{@{}ccc@{}}
    \toprule
    \textbf{ocel\_id [PK] [FK]} & \textbf{ocel\_time} & \textbf{invoice\_inserter} \\
    \midrule
    e5 & 2022-01-14 12:00 & Luke \\
    e6 & 2022-01-16 11:00 & Luke \\
    e9 & 2022-02-02 09:00 & Mario \\
    \bottomrule
  \end{tabular}
  \label{tab:tabInsertInvoice}
\end{table}

\begin{table}[ht]
  \centering
  \caption{[Proposed relational implementation] Event type table: \textbf{event\_SetPaymentBlock}}
  \begin{tabular}{@{}ccc@{}}
    \toprule
    \textbf{ocel\_id [PK] [FK]} & \textbf{ocel\_time} & \textbf{invoice\_blocker} \\
    \midrule
    e11 & 2022-02-03 07:30 & Sam \\
    \bottomrule
  \end{tabular}
  \label{tab:tabSetPaymentBlock}
\end{table}

\begin{table}[ht]
  \centering
  \caption{[Proposed relational implementation] Event type table: \textbf{event\_RemovePaymentBlock}}
  \begin{tabular}{@{}ccc@{}}
    \toprule
    \textbf{ocel\_id [PK] [FK]} & \textbf{ocel\_time} & \textbf{invoice\_block\_rem} \\
    \midrule
    e12 & 2022-02-03 23:30 & Mario \\
    \bottomrule
  \end{tabular}
  \label{tab:tabRemovePaymentBlock}
\end{table}

\begin{table}[ht]
  \centering
  \caption{[Proposed relational implementation] Event type table: \textbf{event\_InsertPayment}}
  \begin{tabular}{@{}ccc@{}}
    \toprule
    \textbf{ocel\_id [PK] [FK]} & \textbf{ocel\_time} & \textbf{payment\_inserter} \\
    \midrule
    e7 & 2022-01-30 23:00 & Robot \\
    e8 & 2022-01-31 22:00 & Robot \\
    e13 & 2022-02-28 23:00 & Robot \\
    \bottomrule
  \end{tabular}
  \label{tab:tabInsertPayment}
\end{table}

\clearpage
\newpage

\subsection{Object Type Tables}
\label{subsec:objTypeTables}

Having defined the table {\bf object} as the general table for the objects (having the identifier of the object as the primary key), we want to define object-type-specific tables for equivalent reasons to the ones explained in Subsection \ref{subsec:eventTypeTables}. Therefore, if $ot \in OT(L)$ is the object type, we have a table called {\bf object\_$\oplus${$map_{OT}$}(ot)}, where $map_{OT}: OT(L) \rightarrow \univ{\Sigma}$ is an injective function\footnote{Which can be the identity function.} mapping the object types to unique identifiers, containing the objects of the given object type along with the history of their attributes. In particular, all the attributes associated with the given object type ($\textrm{oatype}$ in Definition~\ref{def:ocelDef}) are columns of the table, and the values for each object are populated accordingly. The timestamp column here highlights the history of the values of the different attributes. We made the following design choices:
\begin{itemize}
    \item In accordance with the definition of OCEL 2.0, rows possessing the smallest feasible timestamp, specifically \emph{1970-01-01 00:00 UTC} - which equates to $0$ in Definition~\ref{def:ocelDef} - correspond to the initial values of all the attributes for a given object. This uniform timestamp selection, symbolic of the starting point or "epoch", facilitates a consistent reference frame across all objects, aligning seamlessly with the structure proposed by OCEL 2.0. The deliberate choice of this fixed date for the timestamp of $0$ is not arbitrary; instead, it serves a clear purpose by fostering improved compatibility and coherence with the OCEL 2.0 definition.
    \item The rows having a different timestamp have a value in the column \\ \emph{ocel\_changed\_field}, which highlights which other column has been changed.
\end{itemize}
This allows (according to $oaval$ in Definition~\ref{def:ocelDef}) to reconstruct the value of a given attribute at a specified point in time.
For example, given the purchase order {\bf PO1}, the quantity of the order at {\bf 2022-01-11 10:00} is $500$ cows, but it becomes $600$ cows at {\bf 2022-01-13 13:00}.
Also, consider the invoice $R3$, which was blocked and then released.

The object identifier column is pointing (as a foreign key) to the object identifier column of the {\bf object} table.

Examples follow (Table~\ref{tab:tabOtPr}, Table~\ref{tab:tabOtPo}, Table~\ref{tab:tabOtInvoice}, Table~\ref{tab:tabOtPayment}) for all the object types of the running example log.

\newcommand{\tablestyle}{
  \centering
  \renewcommand{\arraystretch}{1.5}
  \setlength{\tabcolsep}{0.5em}
  \small
}

\begin{table}[ht]
  \tablestyle
  \caption{[Proposed relational implementation] Object type table: \textbf{object\_PurchaseRequisition}}
  \begin{tabular}{cccccc}
    \toprule
    \textbf{ocel\_id [FK]} & \textbf{ocel\_time} & \textbf{pr\_product} & \textbf{pr\_quantity} & \textbf{ocel\_changed\_field} \\
    \midrule
    PR1 & 1970-01-01 00:00 UTC & Cows & 500 & ~ \\
    \bottomrule
  \end{tabular}
  \label{tab:tabOtPr}
\end{table}

\begin{table}[ht]
  \tablestyle
  \caption{[Proposed relational implementation] Object type table: \textbf{object\_PurchaseOrder}}
  \begin{tabular}{ccccc}
    \toprule
    \textbf{ocel\_id [FK]} & \textbf{ocel\_time} & \textbf{po\_product} & \textbf{po\_quantity} & \textbf{ocel\_changed\_field} \\
    \midrule
    PO1 & 1970-01-01 00:00 UTC & Cows & 500 & ~ \\
    PO1 & 2022-01-13 12:00 UTC & ~ & 600 & po\_quantity \\
    PO2 & 1970-01-01 01:00 UTC & Notebooks & 1 & ~ \\
    \bottomrule
  \end{tabular}
  \label{tab:tabOtPo}
\end{table}

\begin{table}[ht]
  \tablestyle
  \caption{[Proposed relational implementation] Object type table: \textbf{object\_Invoice}}
  \begin{tabular}{cccc}
    \toprule
    \textbf{ocel\_id [FK]} & \textbf{ocel\_time} & \textbf{is\_blocked} & \textbf{ocel\_changed\_field} \\
    \midrule
    R1 & 1970-01-01 00:00 UTC & No & ~ \\
    R2 & 1970-01-01 00:00 UTC & No & ~ \\
    R3 & 1970-01-01 00:00 UTC & No & ~ \\
    R3 & 2022-02-03 07:30 UTC & Yes & is\_blocked \\
    R3 & 2022-02-03 23:30 UTC & No & is\_blocked \\
    \bottomrule
  \end{tabular}
  \label{tab:tabOtInvoice}
\end{table}

\begin{table}[ht]
  \tablestyle
  \caption{[Proposed relational implementation] Object type table: \textbf{object\_Payment}}
  \begin{tabular}{ccc}
    \toprule
    \textbf{ocel\_id [FK]} & \textbf{ocel\_time} & \textbf{ocel\_changed\_field} \\
    \midrule
    P1 & 1970-01-01 00:00 UTC & ~ \\
    P2 & 1970-01-01 00:00 UTC & ~ \\
    P3 & 1970-01-01 00:00 UTC & ~ \\
    \bottomrule
  \end{tabular}
  \label{tab:tabOtPayment}
\end{table}

\clearpage
\newpage

\subsection{Event-to-Object (E2O) Relationships}
\label{subsec:evToObjRelationships}

The {\bf event\_object} table contains the event-to-object relationships ($\textrm{E2O}$ in Definition~\ref{def:ocelDef}). Therefore, it contains the correlated event identifier and object identifier (foreign key to {\bf event.ocel\_id} and {\bf object.ocel\_id} respectively) with a qualifier explaining the nature of the relationship.
A primary key is set on {\bf event\_object} containing all the columns, therefore ``realizing'' the set proposed in Definition~\ref{def:ocelDef}.
The {\bf event\_object} table of the running example is proposed in Table~\ref{tab:tabE2O}. Note that we can have an event related with different qualifiers to the same object.

\begin{table}[ht]
  \centering
  \caption{[Proposed relational implementation] Table containing the event-to-object (\textbf{event\_object}) relationships}
  \resizebox{\textwidth}{!}{
  \begin{tabular}{ccc}
    \toprule
    \textbf{ocel\_event\_id [PK] [FK]} & \textbf{ocel\_object\_id [PK] [FK]} & \textbf{ocel\_qualifier [PK]} \\
    \midrule
    e1  & PR1 & Regular placement of PR \\
    e2  & PR1 & Regular approval of PR \\
    e3  & PR1 & Created order from PR \\
    e3  & PO1 & Created order with identifier \\
    e4  & PO1 & Change of quantity \\
    e5  & PO1 & Invoice created starting from the PO \\
    e5  & R1  & Invoice created with identifier \\
    e5  & PO1 & Invoice created starting from the PO \\
    e6  & R2  & Invoice created with identifier \\
    e6  & PO1 & Invoice created starting from the PO \\
    e7  & R1  & Payment for the invoice \\
    e7  & P1  & Payment inserted with identifier \\
    e8  & R2  & Payment for the invoice \\
    e8  & P2  & Payment inserted with identifier \\
    e9  & R3  & Invoice created with identifier \\
    e10 & R3  & Purchase order created with maverick buying from \\
    e10 & PO2 & Purchase order created with identifier \\
    e11 & R3  & Payment block due to unethical maverick buying \\
    e12 & R3  & Payment block removed \ldots \\
    e13 & R3  & Payment for the invoice \\
    e13 & P3  & Payment inserted with identifier \\
    \bottomrule
  \end{tabular}
  }
  \label{tab:tabE2O}
\end{table}

\clearpage
\newpage

\subsection{Object-to-Object (O2O) Relationships}
\label{subsec:objObjRelationships}

The {\bf object\_object} table contains the object-to-object relationships ($\textrm{O2O}$ in Definition~\ref{def:ocelDef}). Therefore, it contains the correlated object identifiers (source and target; both are foreign keys to {\bf object.ocel\_id}) with a qualifier explaining the nature of the relationship.
A primary key is set on {\bf object\_object} containing all the columns, therefore ``realizing'' the set proposed in Definition~\ref{def:ocelDef}.
The {\bf object\_object} table of the running example is proposed in Table~\ref{tab:tabO2O}. Note that we can have the same couple of objects related through different qualifiers.

\begin{table}[ht]
  \centering
  \caption{[Proposed relational implementation] Table containing the object-to-object (\textbf{object\_object}) relationships}
  \resizebox{\textwidth}{!}{
  \begin{tabular}{ccc}
    \toprule
    \textbf{ocel\_source\_id [PK] [FK]} & \textbf{ocel\_target\_id [PK] [FK]} & \textbf{ocel\_qualifier [PK]} \\
    \midrule
    PR1 & PO1 & PO from PR \\
    PO1 & R1  & Invoice from PO \\
    PO1 & R2  & Invoice from PO \\
    R1  & P1  & Payment from invoice \\
    R2  & P2  & Payment from invoice \\
    PO2 & R3  & Maverick buying \\
    R3  & P3  & Payment from invoice \\
    \bottomrule
  \end{tabular}
  }
  \label{tab:tabO2O}
\end{table}

\clearpage
\newpage

\subsection{Constraints on the Relational Implementation}
\label{sec:relationalImplementationConstraints}

Clearly, all the elements introduced using the metamodel can be mapped onto the table structures proposed. However, we need to ensure consistency.
Therefore, the following constraints are implemented on the proposed relational implementation:
\begin{itemize}
\item The uniqueness of the event/object types in the tables {\bf event\_map\_type} and \\ {\bf object\_map\_type} is ensured by setting the type as the primary key.
\item The uniqueness of the events/objects in the overall {\bf event} and {\bf object} tables is ensured by setting the identifier as the primary key.
\item The uniqueness of the event identifiers in the specific event type tables is ensured by setting the identifier as the primary key. Since the same objects can appear (by design choice) several times in the specific object type tables, we cannot set the identifier as the primary key in the given setting.
\item There is a foreign key between the specific event type tables and the generic {\bf event} table. This ensures that every identifier appearing in the specific event type tables has been added to the {\bf event} table.
\item There is a foreign key between the specific object type table and the generic {\bf object} table. This ensures that every identifier appearing in the specific object type tables has been added to the {\bf object} table.
\item There is a foreign key between the generic {\bf event} table and {\bf event\_map\_type}, ensuring that every event can be mapped to a specific event type table.
\item There is a foreign key between the generic {\bf object} table and {\bf object\_map\_type}, ensuring that every object is mapped to a specific object type table.
\item The {\bf event\_object} table has two foreign keys, directed towards {\bf event.ocel\_id} and {\bf object.ocel\_id} respectively. This ensures that every identifier appearing in the {\bf event\_object} table is effectively an event or an object. Moreover, the three columns together form the primary key, ensuring that no duplicate rows are contained in the table.
\item The {\bf object\_object} table has two foreign keys, both directed towards {\bf object.ocel\_id}. This ensures that every identifier appearing in the {\bf object\_object} table is effectively an object. Moreover, the three columns together form the primary key, ensuring that no duplicate rows are contained in the table.
\end{itemize}

We can ensure that every event/object is added to the correct event type/object type table by a trigger that checks during the insertion time the \\
{\bf event\_map\_type}/{\bf object\_map\_type} tables to compare the name of the current table with the expected table into which the event/object should be assigned.

We also provide some offline validation constraints as SQL queries. These are available at \href{https://www.ocel-standard.org/2.0/ocel20-schema-relational.pdf}{https://www.ocel-standard.org/2.0/ocel20-schema-relational.pdf}.

\clearpage
\newpage

\section{XML Format}
\label{sec:xmlImplementation}

We propose an XML implementation following Definition~\ref{def:ocelDef}. The timestamps are assumed to follow the ISO format specification \url{https://en.wikipedia.org/wiki/ISO_8601}.

The XML schema is organized as follows. There is a root element with the tag {\bf log}. The log element has the following children:
\begin{itemize}
\item An element with tag {\bf object-types}, containing as many {\bf object-type} elements as types in $OT(L)$. Each {\bf object-type} has a {\bf name} property (which is the object type) and a single child with tag {\bf attributes}:
\begin{itemize}
    \item For every attribute in $\textrm{oatype}(ot)$, the element {\bf attributes} has a child with tag {\bf attribute} and properties {\bf name} (which is the attribute) and {\bf type} (the type of the attribute, which should be considered during the parsing of the values).
\end{itemize}
\item An element with tag {\bf event-types}, containing as many {\bf event-type} elements as types in $OT(L)$. Each {\bf event-type} has a {\bf name} property (which is the event type) and a single child with tag {\bf attributes}:
\begin{itemize}
    \item For every attribute in $\textrm{eatype}(et)$, the element {\bf attributes} has a child with tag {\bf attribute} and properties {\bf name} (which is the attribute) and {\bf type} (the type of the attribute, which should be considered during the parsing of the values).
\end{itemize}
\item An element with tag {\bf events}, containing as many {\bf event} elements as many events are in $E$. An {\bf event} is characterized by:
\begin{itemize}
\item Its properties {\bf id} (the identifier of the event), {\bf type} (the event type of the event, given by \textrm{evtype} in Definition~\ref{def:ocelDef}) and {\bf time} (the timestamp of the event, given by \textrm{time} in Definition~\ref{def:ocelDef}).
\item A child with tag {\bf objects}, containing the related objects to the event (\textrm{E2O} in Definition~\ref{def:ocelDef}; we define the function:
$$\textrm{relobj}(e) = \{ (o, q) ~ \arrowvert ~ (e', q, o) \in \textrm{E2O} ~ \wedge ~ e' = e \}$$
which associates to every event a set of objects along with the qualifier of the relationship). In particular,
for every event-to-object relationship an entry {\bf relobj} is created, having as properties the {\bf object-id} (related object identifier) and {\bf qualifier} (the qualifier of the event-to-object relationship).
\item A child with tag {\bf attributes}, having as many children {\bf attribute} as many attributes are related to the event (the domain of $\textrm{eaval}_{ea}$ in Definition~\ref{def:ocelDef}):
\begin{itemize}
\item Each {\bf attribute} is characterized by a {\bf name} property and the corresponding value is reported as text of the (XML) element.
\end{itemize}
\end{itemize}
\item An element with tag {\bf objects}, containing as many {\bf object} elements as many objects are in $O$. An {\bf object} is characterized by:
\begin{itemize}
    \item Its properties {\bf id} (the identifier of the object) and {\bf type} (the object type of the object, given by \textrm{objtype} in Definition~\ref{def:ocelDef}).
    \item An element with tag {\bf attributes}, containing the different {\bf attribute} of the object.
    \item Each {\bf attribute} is characterized by a {\bf time} property (the timestamp in which the value for the given attribute was recorded), a {\bf name} property, and the corresponding value is reported as the text of the (XML) element. An attribute is valid from the specified {\bf time} (until an attribute with the same name and greater timestamp is recorded).
    \item A child with tag {\bf objects}, containing the related objects to the given object (\textrm{O2O} in Definition~\ref{def:ocelDef}; we define the function:
    $$\textrm{relobj}(o) = \{ (o'', q) ~ \arrowvert ~ (o', q, o'') \in \textrm{O2O} ~ \wedge ~ o' = o \}$$
    which associates to every object a set of objects along with the qualifier of the relationship). In particular,
    for every object-to-object relationship an entry {\bf relobj} is created, having as properties the {\bf object-id} (related object identifier) and {\bf qualifier} (the qualifier of the object-to-object relationship).
\end{itemize}
\end{itemize}

In the remainder of this section, we show an example file and the XSD (XML Schema Definition) that can be used to check consistency.

\subsection{XML Example}
An example (on the running example log) follows.

\lstset{
  basicstyle=\ttfamily\footnotesize,
  columns=fullflexible,
  showstringspaces=false,
  commentstyle=\color{gray}\upshape,
  numbers=left,
  numberstyle=\tiny,
  backgroundcolor=\color{lightgray},
}

\lstdefinelanguage{XML}
{
  morestring=[b]",
  morestring=[s]{>}{<},
  morecomment=[s]{<?}{?>},
  stringstyle=\color{black},
  identifierstyle=\color{darkblue},
  keywordstyle=\color{cyan},
  morekeywords={xmlns,version,type}
}


\begin{lstlisting}[language=XML]
<?xml version='1.0' encoding='UTF-8'?>
<log>
  <object-types>
    <object-type name="Invoice">
      <attributes>
        <attribute name="is_blocked" type="string"/>
      </attributes>
    </object-type>
    <object-type name="Payment">
      <attributes/>
    </object-type>
    <object-type name="Purchase Order">
      <attributes>
        <attribute name="po_product" type="string"/>
        <attribute name="po_quantity" type="string"/>
      </attributes>
    </object-type>
    <object-type name="Purchase Requisition">
      <attributes>
        <attribute name="pr_product" type="string"/>
        <attribute name="pr_quantity" type="string"/>
      </attributes>
    </object-type>
  </object-types>
  <event-types>
    <event-type name="Approve Purchase Requisition">
      <attributes>
        <attribute name="pr_approver" type="string"/>
      </attributes>
    </event-type>
    <event-type name="Change PO Quantity">
      <attributes>
        <attribute name="po_editor" type="string"/>
      </attributes>
    </event-type>
    <event-type name="Create Purchase Order">
      <attributes>
        <attribute name="po_creator" type="string"/>
      </attributes>
    </event-type>
    <event-type name="Create Purchase Requisition">
      <attributes>
        <attribute name="pr_creator" type="string"/>
      </attributes>
    </event-type>
    <event-type name="Insert Invoice">
      <attributes>
        <attribute name="invoice_inserter" type="string"/>
      </attributes>
    </event-type>
    <event-type name="Insert Payment">
      <attributes>
        <attribute name="payment_inserter" type="string"/>
      </attributes>
    </event-type>
    <event-type name="Remove Payment Block">
      <attributes>
        <attribute name="invoice_block_rem" type="string"/>
      </attributes>
    </event-type>
    <event-type name="Set Payment Block">
      <attributes>
        <attribute name="invoice_blocker" type="string"/>
      </attributes>
    </event-type>
  </event-types>
  <objects>
    <object id="R1" type="Invoice">
      <attributes>
        <attribute name="is_blocked" time="1970-01-01T00:00:00Z">No</attribute>
      </attributes>
      <objects>
        <relationship object-id="P1" qualifier="Payment from invoice"/>
      </objects>
    </object>
    <object id="R2" type="Invoice">
      <attributes>
        <attribute name="is_blocked" time="1970-01-01T00:00:00Z">No</attribute>
      </attributes>
      <objects>
        <relationship object-id="P2" qualifier="Payment from invoice"/>
      </objects>
    </object>
    <object id="R3" type="Invoice">
      <attributes>
        <attribute name="is_blocked" time="1970-01-01T00:00:00Z">No</attribute>
        <attribute name="is_blocked" time="2022-02-03T07:30:00Z">Yes</attribute>
        <attribute name="is_blocked" time="2022-02-03T23:30:00Z">No</attribute>
      </attributes>
      <objects>
        <relationship object-id="P3" qualifier="Payment from invoice"/>
      </objects>
    </object>
    <object id="P1" type="Payment">
      <attributes/>
    </object>
    <object id="P2" type="Payment">
      <attributes/>
    </object>
    <object id="P3" type="Payment">
      <attributes/>
    </object>
    <object id="PO1" type="Purchase Order">
      <attributes>
        <attribute name="po_product" time="1970-01-01T00:00:00Z">Cows</attribute>
        <attribute name="po_quantity" time="1970-01-01T00:00:00Z">500</attribute>
        <attribute name="po_quantity" time="2022-01-13T12:00:00Z">600</attribute>
      </attributes>
      <objects>
        <relationship object-id="R1" qualifier="Invoice from PO"/>
        <relationship object-id="R2" qualifier="Invoice from PO"/>
      </objects>
    </object>
    <object id="PO2" type="Purchase Order">
      <attributes>
        <attribute name="po_product" time="1970-01-01T00:00:00Z">Notebooks</attribute>
        <attribute name="po_quantity" time="1970-01-01T00:00:00Z">1</attribute>
      </attributes>
      <objects>
        <relationship object-id="R3" qualifier="Maverick buying"/>
      </objects>
    </object>
    <object id="PR1" type="Purchase Requisition">
      <attributes>
        <attribute name="pr_product" time="1970-01-01T00:00:00Z">Cows</attribute>
        <attribute name="pr_quantity" time="1970-01-01T00:00:00Z">500</attribute>
      </attributes>
      <objects>
        <relationship object-id="PO1" qualifier="PO from PR"/>
      </objects>
    </object>
  </objects>
  <events>
    <event id="e1" type="Create Purchase Requisition" time="2022-01-09T15:00:00Z">
      <attributes>
        <attribute name="pr_creator">Mike</attribute>
      </attributes>
      <objects>
        <relationship object-id="PR1" qualifier="Regular placement of PR"/>
      </objects>
    </event>
    <event id="e2" type="Approve Purchase Requisition" time="2022-01-09T16:30:00Z">
      <attributes>
        <attribute name="pr_approver">Tania</attribute>
      </attributes>
      <objects>
        <relationship object-id="PR1" qualifier="Regular approval of PR"/>
      </objects>
    </event>
    <event id="e3" type="Create Purchase Order" time="2022-01-10T09:15:00Z">
      <attributes>
        <attribute name="po_creator">Mike</attribute>
      </attributes>
      <objects>
        <relationship object-id="PR1" qualifier="Created order from PR"/>
        <relationship object-id="PO1" qualifier="Created order with identifier"/>
      </objects>
    </event>
    <event id="e4" type="Change PO Quantity" time="2022-01-13T12:00:00Z">
      <attributes>
        <attribute name="po_editor">Mike</attribute>
      </attributes>
      <objects>
        <relationship object-id="PO1" qualifier="Change of quantity"/>
      </objects>
    </event>
    <event id="e5" type="Insert Invoice" time="2022-01-14T12:00:00Z">
      <attributes>
        <attribute name="invoice_inserter">Luke</attribute>
      </attributes>
      <objects>
        <relationship object-id="PO1" qualifier="Invoice created starting from the PO"/>
        <relationship object-id="R1" qualifier="Invoice created with identifier"/>
      </objects>
    </event>
    <event id="e6" type="Insert Invoice" time="2022-01-16T11:00:00Z">
      <attributes>
        <attribute name="invoice_inserter">Luke</attribute>
      </attributes>
      <objects>
        <relationship object-id="PO1" qualifier="Invoice created starting from the PO"/>
        <relationship object-id="R2" qualifier="Invoice created with identifier"/>
      </objects>
    </event>
    <event id="e7" type="Insert Payment" time="2022-01-30T23:00:00Z">
      <attributes>
        <attribute name="payment_inserter">Robot</attribute>
      </attributes>
      <objects>
        <relationship object-id="R1" qualifier="Payment for the invoice"/>
        <relationship object-id="P1" qualifier="Payment inserted with identifier"/>
      </objects>
    </event>
    <event id="e8" type="Insert Payment" time="2022-01-31T22:00:00Z">
      <attributes>
        <attribute name="payment_inserter">Robot</attribute>
      </attributes>
      <objects>
        <relationship object-id="R2" qualifier="Payment for the invoice"/>
        <relationship object-id="P2" qualifier="Payment created with identifier"/>
      </objects>
    </event>
    <event id="e9" type="Insert Invoice" time="2022-02-02T09:00:00Z">
      <attributes>
        <attribute name="invoice_inserter">Mario</attribute>
      </attributes>
      <objects>
        <relationship object-id="R3" qualifier="Invoice created with identifier"/>
      </objects>
    </event>
    <event id="e10" type="Create Purchase Order" time="2022-02-02T17:00:00Z">
      <attributes>
        <attribute name="po_creator">Mario</attribute>
      </attributes>
      <objects>
        <relationship object-id="R3" qualifier="Purchase order created with maverick buying from"/>
        <relationship object-id="PO2" qualifier="Purchase order created with identifier"/>
      </objects>
    </event>
    <event id="e11" type="Set Payment Block" time="2022-02-03T07:30:00Z">
      <attributes>
        <attribute name="invoice_blocker">Mario</attribute>
      </attributes>
      <objects>
        <relationship object-id="R3" qualifier="Payment block due to unethical maverick buying"/>
      </objects>
    </event>
    <event id="e12" type="Remove Payment Block" time="2022-02-03T23:30:00Z">
      <attributes>
        <attribute name="invoice_block_rem">Mario</attribute>
      </attributes>
      <objects>
        <relationship object-id="R3" qualifier="Payment block removed ..."/>
      </objects>
    </event>
    <event id="e13" type="Insert Payment" time="2022-02-28T23:00:00Z">
      <attributes>
        <attribute name="payment_inserter">Robot</attribute>
      </attributes>
      <objects>
        <relationship object-id="R3" qualifier="Payment for the invoice"/>
        <relationship object-id="P3" qualifier="Payment inserted with identifier"/>
      </objects>
    </event>
  </events>
</log>
\end{lstlisting}

\subsection{XML Schema Definition}
A machine-readable XML Schema Definition (XSD) file is provided to check whether an example XML OCEL 2.0 is valid, see \href{https://www.ocel-standard.org/2.0/ocel20-schema-xml.xsd}{https://www.ocel-standard.org/2.0/ocel20-schema-xml.xsd} Numerous tools are available to validate an XML file against an XSD file.

\begin{lstlisting}[language=XML]
<xs:schema attributeFormDefault="unqualified" elementFormDefault="qualified"
  xmlns:xs="http://www.w3.org/2001/XMLSchema">
  <xs:element name="attribute">
    <xs:complexType>
      <xs:simpleContent>
        <xs:extension base="xs:string">
          <xs:attribute type="xs:string" name="name" use="required"/>
          <xs:attribute name="type" use="optional">
            <xs:simpleType>
              <xs:restriction base="xs:string">
                <xs:enumeration value="string"/>
                <xs:enumeration value="time"/>
                <xs:enumeration value="integer"/>
                <xs:enumeration value="float"/>
                <xs:enumeration value="boolean"/>
              </xs:restriction>
            </xs:simpleType>
          </xs:attribute>
          <xs:attribute type="xs:dateTime" name="time" use="optional"/>
        </xs:extension>
      </xs:simpleContent>
    </xs:complexType>
  </xs:element>
  <xs:element name="attributes">
    <xs:complexType mixed="true">
      <xs:sequence>
        <xs:element ref="attribute" maxOccurs="unbounded" minOccurs="0"/>
      </xs:sequence>
    </xs:complexType>
  </xs:element>
  <xs:element name="object-type">
    <xs:complexType>
      <xs:sequence>
        <xs:element ref="attributes"/>
      </xs:sequence>
      <xs:attribute type="xs:string" name="name" use="required"/>
    </xs:complexType>
    <xs:key name="objectTypeAttributeKey">
      <xs:selector xpath="attributes/attribute"/>
      <xs:field xpath="@name"/>
    </xs:key>
  </xs:element>
  <xs:element name="event-type">
    <xs:complexType>
      <xs:sequence>
        <xs:element ref="attributes"/>
      </xs:sequence>
      <xs:attribute type="xs:string" name="name" use="required"/>
    </xs:complexType>
    <xs:key name="eventTypeAttributeKey">
      <xs:selector xpath="attributes/attribute"/>
      <xs:field xpath="@name"/>
    </xs:key>
  </xs:element>
  <xs:element name="object">
    <xs:complexType mixed="true">
      <xs:sequence>
        <xs:element ref="attributes" minOccurs="0"/>
        <xs:element ref="objects" minOccurs="0"/>
      </xs:sequence>
      <xs:attribute type="xs:string" name="object-id" use="optional"/>
      <xs:attribute type="xs:string" name="qualifier" use="optional"/>
      <xs:attribute type="xs:string" name="id" use="optional"/>
      <xs:attribute type="xs:string" name="type" use="required"/>
    </xs:complexType>
  </xs:element>
  <xs:element name="objects">
    <xs:complexType>
      <xs:sequence>
        <xs:element ref="object" maxOccurs="unbounded" minOccurs="0"/>
      </xs:sequence>
    </xs:complexType>
  </xs:element>
  <xs:element name="event">
    <xs:complexType>
      <xs:sequence>
        <xs:element ref="attributes"/>
        <xs:element ref="objects"/>
      </xs:sequence>
      <xs:attribute type="xs:string" name="id" use="required"/>
      <xs:attribute type="xs:string" name="type" use="required"/>
      <xs:attribute type="xs:dateTime" name="time" use="required"/>
    </xs:complexType>
  </xs:element>
  <xs:element name="object-types">
    <xs:complexType>
      <xs:sequence>
        <xs:element ref="object-type" maxOccurs="unbounded" minOccurs="0"/>
      </xs:sequence>
    </xs:complexType>
  </xs:element>
  <xs:element name="event-types">
    <xs:complexType>
      <xs:sequence>
        <xs:element ref="event-type" maxOccurs="unbounded" minOccurs="0"/>
      </xs:sequence>
    </xs:complexType>
  </xs:element>
  <xs:element name="events">
    <xs:complexType>
      <xs:sequence>
        <xs:element ref="event" maxOccurs="unbounded" minOccurs="0"/>
      </xs:sequence>
    </xs:complexType>
  </xs:element>
  <xs:element name="log">
    <xs:complexType>
      <xs:sequence>
        <xs:element ref="object-types"/>
        <xs:element ref="event-types"/>
        <xs:element ref="objects"/>
        <xs:element ref="events"/>
      </xs:sequence>
    </xs:complexType>
    <xs:key name="objectTypeKey">
      <xs:selector xpath="object-types/object-type"/>
      <xs:field xpath="@name"/>
    </xs:key>
    <xs:key name="objectKey">
      <xs:selector xpath="objects/object"/>
      <xs:field xpath="@id"/>
    </xs:key>
    <xs:key name="eventTypeKey">
      <xs:selector xpath="event-types/event-type"/>
      <xs:field xpath="@name"/>
    </xs:key>
    <xs:key name="eventKey">
      <xs:selector xpath="events/event"/>
      <xs:field xpath="@id"/>
    </xs:key>
    <xs:keyref name="objectTypeRef" refer="objectTypeKey">
      <xs:selector xpath="objects/object"/>
      <xs:field xpath="@type"/>
    </xs:keyref>
    <xs:keyref name="eventTypeRef" refer="eventTypeKey">
      <xs:selector xpath="events/event"/>
      <xs:field xpath="@type"/>
    </xs:keyref>
    <xs:keyref name="objectObjectRef" refer="objectKey">
      <xs:selector xpath="objects/object/objects/object"/>
      <xs:field xpath="@object-id"/>
    </xs:keyref>
    <xs:keyref name="eventObjectRef" refer="objectKey">
      <xs:selector xpath="events/event/objects/object"/>
      <xs:field xpath="@object-id"/>
    </xs:keyref>
    <xs:keyref name="objectAttributeRef" refer="objectTypeAttributeKey">
      <xs:selector xpath="objects/object/attributes/attribute"/>
      <xs:field xpath="@name"/>
    </xs:keyref>
    <xs:keyref name="eventAttributeRef" refer="eventTypeAttributeKey">
      <xs:selector xpath="events/event/attributes/attribute"/>
      <xs:field xpath="@name"/>
    </xs:keyref>
  </xs:element>
</xs:schema>
\end{lstlisting}

\clearpage
\newpage

\section{JSON Format}
\label{sec:jsonImplementation}

The JSON format provides a lightweight structure for web-native process mining applications.
It is conceptually similar to the XML format with its top-level arrays \texttt{events}, \texttt{eventTypes}, \texttt{objects}, and \texttt{objectTypes}.

In the following, we describe these four top-level properties in detail.

\begin{itemize}
\item The top-level \texttt{event} array contains event objects with the properties \texttt{id}, \texttt{type} (referencing the name of an event type), and \texttt{time} (ISO format). An event's \texttt{attributes} are structured into an array of attribute objects with \texttt{name} and \texttt{value} properties.
The event's event-to-object relationships are listed in the \textit{relationships} array with \texttt{objectId} and \texttt{qualifier}.

\item The top-level \texttt{eventTypes} array contains event type objects with a \texttt{name} and a list of attributes with \texttt{name} and \texttt{value} properties. Valid types are \textbf{string}, \textbf{time}, \textbf{integer}, \textbf{float}, and \textbf{boolean}.

\item The top-level \texttt{object} array contains a list of objects as JSON object, with properties \texttt{id} and \texttt{type} (referencing the name of an object type). The attributes property contains an array of attributes with the properties \texttt{name}, \texttt{time} (ISO format), and \texttt{value}.

\item Finally, the top-level objectTypes array contains object type description objects with a \texttt{name} and a list of attributes with \texttt{name} and \texttt{value} properties. Valid types are \textbf{string}, \textbf{time}, \textbf{integer}, \textbf{float}, and \textbf{boolean}.
\end{itemize}

\subsection{JSON Example}
As an example, we show the running example formatted as a JSON document.

\colorlet{punct}{red!60!black}
\definecolor{background}{HTML}{EEEEEE}
\definecolor{delim}{RGB}{20,105,176}
\colorlet{numb}{magenta!60!black}
\definecolor{mymauve}{rgb}{0.58, 0, 0.82}
\definecolor{mygray}{rgb}{0.5, 0.5, 0.5}

\lstdefinelanguage{json}{
  basicstyle=\normalfont\ttfamily\footnotesize,
  numbers=left,
  numberstyle=\scriptsize,
  stepnumber=1,
  numbersep=8pt,
  showstringspaces=false,
  breaklines=true,
  frame=lines,
  backgroundcolor=\color{lightgray},
  stringstyle=\color{mymauve},
  keywordstyle=\color{blue},
  numberstyle=\color{mygray},
  commentstyle=\color{mygray},
  morecomment=[l]{//}, 
  morecomment=[s]{/*}{*/},
  morecomment=[l]{///},
  morestring=[b]",
  morekeywords={true, false, null}
}

\begin{lstlisting}[language=json]
{
  "objectTypes": [
    {
      "name": "Invoice",
      "attributes": [
        {
          "name": "is_blocked",
          "type": "string"
        }
      ]
    },
    {
      "name": "Payment",
      "attributes": []
    },
    {
      "name": "Purchase Order",
      "attributes": [
        {
          "name": "po_product",
          "type": "string"
        },
        {
          "name": "po_quantity",
          "type": "string"
        }
      ]
    },
    {
      "name": "Purchase Requisition",
      "attributes": [
        {
          "name": "pr_product",
          "type": "string"
        },
        {
          "name": "pr_quantity",
          "type": "string"
        }
      ]
    }
  ],
  "eventTypes": [
    {
      "name": "Approve Purchase Requisition",
      "attributes": [
        {
          "name": "pr_approver",
          "type": "string"
        }
      ]
    },
    {
      "name": "Change PO Quantity",
      "attributes": [
        {
          "name": "po_editor",
          "type": "string"
        }
      ]
    },
    {
      "name": "Create Purchase Order",
      "attributes": [
        {
          "name": "po_creator",
          "type": "string"
        }
      ]
    },
    {
      "name": "Create Purchase Requisition",
      "attributes": [
        {
          "name": "pr_creator",
          "type": "string"
        }
      ]
    },
    {
      "name": "Insert Invoice",
      "attributes": [
        {
          "name": "invoice_inserter",
          "type": "string"
        }
      ]
    },
    {
      "name": "Insert Payment",
      "attributes": [
        {
          "name": "payment_inserter",
          "type": "string"
        }
      ]
    },
    {
      "name": "Remove Payment Block",
      "attributes": [
        {
          "name": "invoice_block_rem",
          "type": "string"
        }
      ]
    },
    {
      "name": "Set Payment Block",
      "attributes": [
        {
          "name": "invoice_blocker",
          "type": "string"
        }
      ]
    }
  ],
  "objects": [
    {
      "id": "R1",
      "type": "Invoice",
      "attributes": [
        {
          "name": "is_blocked",
          "time": "1970-01-01T00:00:00Z",
          "value": "No"
        }
      ],
      "relationships": [
        {
          "objectId": "P1",
          "qualifier": "Payment from invoice"
        }
      ]
    },
    {
      "id": "R2",
      "type": "Invoice",
      "attributes": [
        {
          "name": "is_blocked",
          "time": "1970-01-01T00:00:00Z",
          "value": "No"
        }
      ],
      "relationships": [
        {
          "objectId": "P2",
          "qualifier": "Payment from invoice"
        }
      ]
    },
    {
      "id": "R3",
      "type": "Invoice",
      "attributes": [
        {
          "name": "is_blocked",
          "time": "1970-01-01T00:00:00Z",
          "value": "No"
        },
        {
          "name": "is_blocked",
          "time": "2022-02-03T07:30:00Z",
          "value": "Yes"
        },
        {
          "name": "is_blocked",
          "time": "2022-02-03T23:30:00Z",
          "value": "No"
        }
      ],
      "relationships": [
        {
          "objectId": "P3",
          "qualifier": "Payment from invoice"
        }
      ]
    },
    {
      "id": "P1",
      "type": "Payment"
    },
    {
      "id": "P2",
      "type": "Payment"
    },
    {
      "id": "P3",
      "type": "Payment"
    },
    {
      "id": "PO1",
      "type": "Purchase Order",
      "attributes": [
        {
          "name": "po_product",
          "time": "1970-01-01T00:00:00Z",
          "value": "Cows"
        },
        {
          "name": "po_quantity",
          "time": "1970-01-01T00:00:00Z",
          "value": "500"
        },
        {
          "name": "po_quantity",
          "time": "2022-01-13T12:00:00Z",
          "value": "600"
        }
      ],
      "relationships": [
        {
          "objectId": "R1",
          "qualifier": "Invoice from PO"
        },
        {
          "objectId": "R2",
          "qualifier": "Invoice from PO"
        }
      ]
    },
    {
      "id": "PO2",
      "type": "Purchase Order",
      "attributes": [
        {
          "name": "po_product",
          "time": "1970-01-01T00:00:00Z",
          "value": "Notebooks"
        },
        {
          "name": "po_quantity",
          "time": "1970-01-01T00:00:00Z",
          "value": "1"
        }
      ],
      "relationships": [
        {
          "objectId": "R3",
          "qualifier": "Maverick buying"
        }
      ]
    },
    {
      "id": "PR1",
      "type": "Purchase Requisition",
      "attributes": [
        {
          "name": "pr_product",
          "time": "1970-01-01T00:00:00Z",
          "value": "Cows"
        },
        {
          "name": "pr_quantity",
          "time": "1970-01-01T00:00:00Z",
          "value": "500"
        }
      ],
      "relationships": [
        {
          "objectId": "PO1",
          "qualifier": "PO from PR"
        }
      ]
    }
  ],
  "events": [
    {
      "id": "e1",
      "type": "Create Purchase Requisition",
      "time": "2022-01-09T15:00:00Z",
      "attributes": [
        {
          "name": "pr_creator",
          "value": "Mike"
        }
      ],
      "relationships": [
        {
          "objectId": "PR1",
          "qualifier": "Regular placement of PR"
        }
      ]
    },
    {
      "id": "e2",
      "type": "Approve Purchase Requisition",
      "time": "2022-01-09T16:30:00Z",
      "attributes": [
        {
          "name": "pr_approver",
          "value": "Tania"
        }
      ],
      "relationships": [
        {
          "objectId": "PR1",
          "qualifier": "Regular approval of PR"
        }
      ]
    },
    {
      "id": "e3",
      "type": "Create Purchase Order",
      "time": "2022-01-10T09:15:00Z",
      "attributes": [
        {
          "name": "po_creator",
          "value": "Mike"
        }
      ],
      "relationships": [
        {
          "objectId": "PR1",
          "qualifier": "Created order from PR"
        },
        {
          "objectId": "PO1",
          "qualifier": "Created order with identifier"
        }
      ]
    },
    {
      "id": "e4",
      "type": "Change PO Quantity",
      "time": "2022-01-13T12:00:00Z",
      "attributes": [
        {
          "name": "po_editor",
          "value": "Mike"
        }
      ],
      "relationships": [
        {
          "objectId": "PO1",
          "qualifier": "Change of quantity"
        }
      ]
    },
    {
      "id": "e5",
      "type": "Insert Invoice",
      "time": "2022-01-14T12:00:00Z",
      "attributes": [
        {
          "name": "invoice_inserter",
          "value": "Luke"
        }
      ],
      "relationships": [
        {
          "objectId": "PO1",
          "qualifier": "Invoice created starting from the PO"
        },
        {
          "objectId": "R1",
          "qualifier": "Invoice created with identifier"
        }
      ]
    },
    {
      "id": "e6",
      "type": "Insert Invoice",
      "time": "2022-01-16T11:00:00Z",
      "attributes": [
        {
          "name": "invoice_inserter",
          "value": "Luke"
        }
      ],
      "relationships": [
        {
          "objectId": "PO1",
          "qualifier": "Invoice created starting from the PO"
        },
        {
          "objectId": "R2",
          "qualifier": "Invoice created with identifier"
        }
      ]
    },
    {
      "id": "e7",
      "type": "Insert Payment",
      "time": "2022-01-30T23:00:00Z",
      "attributes": [
        {
          "name": "payment_inserter",
          "value": "Robot"
        }
      ],
      "relationships": [
        {
          "objectId": "R1",
          "qualifier": "Payment for the invoice"
        },
        {
          "objectId": "P1",
          "qualifier": "Payment inserted with identifier"
        }
      ]
    },
    {
      "id": "e8",
      "type": "Insert Payment",
      "time": "2022-01-31T22:00:00Z",
      "attributes": [
        {
          "name": "payment_inserter",
          "value": "Robot"
        }
      ],
      "relationships": [
        {
          "objectId": "R2",
          "qualifier": "Payment for the invoice"
        },
        {
          "objectId": "P2",
          "qualifier": "Payment created with identifier"
        }
      ]
    },
    {
      "id": "e9",
      "type": "Insert Invoice",
      "time": "2022-02-02T09:00:00Z",
      "attributes": [
        {
          "name": "invoice_inserter",
          "value": "Mario"
        }
      ],
      "relationships": [
        {
          "objectId": "R3",
          "qualifier": "Invoice created with identifier"
        }
      ]
    },
    {
      "id": "e10",
      "type": "Create Purchase Order",
      "time": "2022-02-02T17:00:00Z",
      "attributes": [
        {
          "name": "po_creator",
          "value": "Mario"
        }
      ],
      "relationships": [
        {
          "objectId": "R3",
          "qualifier": "Purchase order created with maverick buying from"
        },
        {
          "objectId": "PO2",
          "qualifier": "Purchase order created with identifier"
        }
      ]
    },
    {
      "id": "e11",
      "type": "Set Payment Block",
      "time": "2022-02-03T07:30:00Z",
      "attributes": [
        {
          "name": "invoice_blocker",
          "value": "Mario"
        }
      ],
      "relationships": [
        {
          "objectId": "R3",
          "qualifier": "Payment block due to unethical maverick buying"
        }
      ]
    },
    {
      "id": "e12",
      "type": "Remove Payment Block",
      "time": "2022-02-03T23:30:00Z",
      "attributes": [
        {
          "name": "invoice_block_rem",
          "value": "Mario"
        }
      ],
      "relationships": [
        {
          "objectId": "R3",
          "qualifier": "Payment block removed ..."
        }
      ]
    },
    {
      "id": "e13",
      "type": "Insert Payment",
      "time": "2022-02-28T23:00:00Z",
      "attributes": [
        {
          "name": "payment_inserter",
          "value": "Robot"
        }
      ],
      "relationships": [
        {
          "objectId": "R3",
          "qualifier": "Payment for the invoice"
        },
        {
          "objectId": "P3",
          "qualifier": "Payment inserted with identifier"
        }
      ]
    }
  ]
}
\end{lstlisting}

\subsection{JSON Schema Definition}

We defined a validation schema for the OCEL 2.0 JSON specification. 
The schema is reported in the following snippet and can be downloaded from 
\href{https://www.ocel-standard.org/2.0/ocel20-schema-json.json}{https://www.ocel-standard.org/2.0/ocel20-schema-json.json}.

\begin{lstlisting}[language=json]
{
    "$schema": "http://json-schema.org/draft-07/schema#",
    "type": "object",
    "properties": {
        "eventTypes": {
            "type": "array",
            "items": {
                "type": "object",
                "properties": {
                    "name": { "type": "string" },
                    "attributes": {
                        "type": "array",
                        "items": {
                            "type": "object",
                            "properties": {
                                "name": { "type": "string" },
                                "type": { "type": "string" }
                            },
                            "required": ["name", "type"]
                        }
                    }
                },
                "required": ["name", "attributes"]
            }
        },
        "objectTypes": {
            "type": "array",
            "items": {
                "type": "object",
                "properties": {
                    "name": { "type": "string" },
                    "attributes": {
                        "type": "array",
                        "items": {
                            "type": "object",
                            "properties": {
                                "name": { "type": "string" },
                                "type": { "type": "string" }
                            },
                            "required": ["name", "type"]
                        }
                    }
                },
                "required": ["name", "attributes"]
            }
        },
        "events": {
            "type": "array",
            "items": {
                "type": "object",
                "properties": {
                    "id": { "type": "string" },
                    "type": { "type": "string" },
                    "time": { "type": "string", "format": "date-time" },
                    "attributes": {
                        "type": "array",
                        "items": {
                            "type": "object",
                            "properties": {
                                "name": { "type": "string" },
                                "value": { "type": "string" }
                            },
                            "required": ["name", "value"]
                        }
                    },
                    "relationships": {
                        "type": "array",
                        "items": {
                            "type": "object",
                            "properties": {
                                "objectId": { "type": "string" },
                                "qualifier": { "type": "string" }
                            },
                            "required": ["objectId", "qualifier"]
                        }
                    }
                },
                "required": ["id", "type", "time"]
            }
        },
        "objects": {
            "type": "array",
            "items": {
                "type": "object",
                "properties": {
                    "id": { "type": "string" },
                    "type": { "type": "string" },
                    "relationships": {
                        "type": "array",
                        "items": {
                            "type": "object",
                            "properties": {
                                "objectId": { "type": "string" },
                                "qualifier": { "type": "string" }
                            },
                            "required": ["objectId", "qualifier"]
                        }
                    },
                    "attributes": {
                        "type": "array",
                        "items": {
                            "type": "object",
                            "properties": {
                                "name": { "type": "string" },
                                "value": { "type": "string" },
                                "time": { "type": "string", "format": "date-time" }
                            },
                            "required": ["name", "value", "time"]
                        }
                    }
                },
                "required": ["id", "type"]
            }
        }
    },
    "required": ["eventTypes", "objectTypes", "events", "objects"]
}
\end{lstlisting}

\clearpage
\newpage

\section{Conclusion}
\label{sec:conclusion}

This document provides a comprehensive introduction to the OCEL 2.0 standard. 
Given the increasing importance of Object-Centric Process Mining (OCPM),
it is important to be able to standardize Object-Centric Event Data (OCED).
OCEL 2.0 aims to provide a middle ground between simplicity and expressiveness, building upon experiences with OCEL 1.0 over the past three years.
We first provided a contextual understanding of the object-centric process mining landscape and then discussed the motivation and the necessities that led to the creation of the OCEL 2.0 standard.

Why is this relevant?
\begin{itemize}
\item Using OCEL 2.0, it is possible to create a system-agnostic, 
single source of truth. Event data should capture real business-relevant events without being limited by a single-case notion.
\item  We no longer need to create a new event log for each process (or view on a selected process). Using traditional event logs, 
there may be overlapping logs that refer to products, suppliers, etc. 
This leads to duplication and inconsistencies. 
Using OCEL 2.0, views can be created on
demand without going back to the source systems.
\item OCEL 2.0 and the OCPM techniques that build upon it allow us to
stay closer to reality, also allowing organizations to uncover problems that
live at the intersection points of processes and organizational units. 
\end{itemize}

Through a detailed presentation of formal definitions, we built the mathematical foundation that equips readers to effectively utilize and understand the OCEL 2.0 standard in a scientific context.
Our illustrative running example served to bridge the gap between abstract theory and practical application, enabling readers to fully grasp how the principles of OCEL 2.0 come into play.

The detailed overview of the relational SQLite, XML, and JSON implementations demonstrated the versatility and compatibility of the standard across multiple technological contexts. Together, these make OCEL 2.0 an accessible and practical choice for both academics and industry practitioners.

Compared to OCEL 1.0, OCEL 2.0 allows for changes in objects, is able to relate objects directly, and can qualify relationships between objects and events.
We hope that this will fuel new OCPM techniques using these extensions.
The relational SQLite implementation also shows that this standard goes beyond 
file formats like XML, JSON, and Excel. Because the reference storage format for the XES standard was XML, people often misunderstood the XES standard. The XES metamodel can be implemented in many different ways. 
However, some vendors used the bulkiness XML as an excuse not to support XES, 
thus blocking any form of standardization.
However, the critical point is the standardization and unification of concepts (not the serialization of data using a specific syntax). 
Therefore, we encourage researchers and software companies to come up with new storage formats implementing the OCEL 2.0 metamodel and provide conversations.
The SQLite, XML, and JSON formats are just examples, and we provide tools to convert any one of them into the two other formats. However, we encourage developers to create novel, highly scalable formats. 
This is one of the reasons why we kept OCEL 2.0 as simple and as concrete as possible. 

We also hope that OCEL 2.0 will also be the basis for creating \textit{standard object and event types} for different application domains. 
Organizations struggle to use ontologies and related technologies because the added value of extensively modeling data is not so clear. 
OCPM can address this problem. Event data in OCEL 2.0 format enables process discovery, conformance checking, performance analysis, and operational support without the need to process the data further. 
However, any organization has standard processes like Order-to-Cash (O2C) and Procure-to-Pay (P2P) that talk about suppliers, customers, orders, items, shipments, etc. There is no need to reinvent the wheel. Also, one would like to keep these things system-agnostic. 
Definitions of object types and event types and their attributes need to be standardized. It is possible to define taxonomies of object types and event types using inheritance notions \cite{mathematics-OCPM-wvda-2023}. This creates possibilities for both generative and discriminative Artificial Intelligence (AI). Therefore, researchers and solution providers should focus on creating standard object and event types and the corresponding normative object-centric process models. This will prevent organizations from starting from scratch when applying process mining and AI for the first time.

Lastly, it is essential to note that the rapidly evolving field of object-centric process mining continues to present new challenges and opportunities. As such, the OCEL 2.0 standard, despite its substantial contribution, should be seen as a stepping stone in this exciting journey rather than a final destination.
Furthermore, this standard is intended to help pave the path for the development of process mining techniques supporting the journey to more sustainable operational practices. We encourage further research and development efforts to build on this foundation, with the aim of continuously advancing the field to new heights of innovation and practical value.

You can find further details of OCEL 2.0, example event logs, and tool support, on our homepage:

\begin{center}
  {\large \url{https://www.ocel-standard.org}}
\end{center}

\newpage
\clearpage

\section*{Acknowledgments}
Funded by the Deutsche Forschungsgemeinschaft (DFG, German Research Foundation) under Germany's Excellence Strategy - EXC-2023 Internet of Production - 390621612.
We also thank the Alexander von Humboldt (AvH) Stiftung for supporting our research.

\newpage
\clearpage

\bibliography{bibliography}

\begin{thebibliography}{10}

\bibitem{process-mining-book-2016}
{W.M.P. van der} Aalst.
\newblock {\em {Process Mining: Data Science in Action}}.
\newblock Springer-Verlag, Berlin, 2016.

\bibitem{mathematics-OCPM-wvda-2023}
{W.M.P van der} Aalst.
\newblock {Object-Centric Process Mining: Unraveling the Fabric of Real
  Processes}.
\newblock {\em Mathematics}, 11(12):2691, 2023.

\bibitem{ocpn_fi_2020}
{W.M.P. van der} Aalst and A.~Berti.
\newblock {Discovering Object-Centric Petri Nets}.
\newblock {\em Fundamenta Informaticae}, 175(1-4):1--40, 2020.

\bibitem{PMhandbook-SS22}
{W.M.P. van der} Aalst and J.~Carmona, editors.
\newblock {\em Process Mining Handbook}, volume 448 of {\em Lecture Notes in
  Business Information Processing}.
\newblock Springer-Verlag, Berlin, 2022.

\bibitem{ieee-CIM-XES-2017}
G.~Acampora, A.~Vitiello, {B. Di} Stefano, {W. van der} Aalst, C.~G{\"u}nther,
  and E.~Verbeek.
\newblock {IEEE 1849: The XES Standard - The Second IEEE Standard Sponsored by
  IEEE Computational Intelligence Society}.
\newblock {\em IEEE Computational Intelligence Magazine}, 12(2):4--8, 2017.

\bibitem{niklas-ocel-framework-icsoc-2022}
J.N. Adams, G.~Park, S.~Levich, D.~Schuster, and {W.M.P. van der} Aalst.
\newblock {A Framework for Extracting and Encoding Features from Object-Centric
  Event Data}.
\newblock In J.~Troya, B.Medjahed, M.Piattini, L.~Yao, P.~Fern{\'{a}}ndez, and
  A.~Ruiz{-}Cort{\'{e}}s, editors, {\em {International Conference on
  Service-Oriented Computing (ICSOC 2022)}}, volume 13740 of {\em Lecture Notes
  in Computer Science}, pages 36--53. Springer-Verlag, Berlin, 2022.

\bibitem{OCEL-standard-2020-people}
A.F. Ghahfarokhi, G.~Park, A.~Berti, and {W.M.P. van der} Aalst.
\newblock {OCEL Standard}.
\newblock www.ocel-standard.org, 2020.

\bibitem{OCEL-ADBIS2021}
A.F. Ghahfarokhi, G.~Park, A.~Berti, and {W.M.P. van der} Aalst.
\newblock {OCEL: A Standard for Object-Centric Event Logs}.
\newblock In L.~Bellatreche, M.~Dumas, and P.~Karras, editors, {\em New Trends
  in Database and Information Systems (Short Papers ADBIS 2021)}, volume 1450
  of {\em {Communications in Computer and Information Science}}, pages
  169--175. Springer-Verlag, Berlin, 2021.

\bibitem{XES-standard-2023}
{IEEE}.
\newblock {IEEE Standard for eXtensible Event Stream (XES) for Achieving
  Interoperability in Event Logs and Event Streams}.
\newblock {\em {IEEE Std 1849-2023 (Revision of IEEE Std 1849-2016)}}, pages
  1--55, 2023.
\newblock 10.1109/IEEESTD.2023.10267858.

\bibitem{XES-standard-2010}
{IEEE Task Force on Process Mining}.
\newblock {XES Standard Definition}.
\newblock www.xes-standard.org, 2010.

\bibitem{gartner-MQ-PM-2023}
M.~Kerremans, K.~Iijima, A.~Sachelarescu, N.~Duffy, and D.~Sugden.
\newblock {Magic Quadrant for Process Mining Tools, Gartner Research Note
  GG00774746}.
\newblock www.gartner.com, 2023.

\bibitem{XES-survey-icpm-ws-lnbip-2021}
M.T. Wynn, J.~Lebherz, {W.M.P. van der} Aalst, R.~Accorsi, {C. Di} Ciccio,
  L.~Jayarathna, and H.M.W. Verbeek.
\newblock {Rethinking the Input for Process Mining: Insights from the XES
  Survey and Workshop}.
\newblock In J.~Munoz-Gama and X.~Lu, editors, {\em {Process Mining Workshops
  of the International Conference on Process Mining (Revised Selected
  Papers)}}, volume 433 of {\em Lecture Notes in Business Information
  Processing}, pages 3--16. Springer-Verlag, Berlin, 2021.

\end{thebibliography}

\bibliographystyle{plain}
\end{document}